\documentclass[12pt]{article}
\usepackage{cite}
\usepackage{epsfig,epsf}
\usepackage{times} 
 \usepackage{longtable}
 \usepackage[nottoc]{tocbibind}
 \usepackage{booktabs,caption,morefloats}
\usepackage[flushleft]{threeparttable}
\usepackage{setspace}
\usepackage{graphicx}
\usepackage{subcaption}
\begin{document}
\parindent 00pt
{\LARGE{\bf{Orientation and Alignment Dynamics of Polar Molecule Driven by Shaped Laser Pulses}}}\\[9mm]
Vijit V. Nautiyal$^{a}$, Sumana Devi$^{a,b}$, Ashish Tyagi$^{c}$, Bhawna Vidhani$^{d}$, Anjali Maan$^{e}$, Vinod Prasad$^{*c}$\let\thefootnote\relax\footnotetext{$^*$Corresponding author,{E-mail:vprasad@ss.du.ac.in}}
\\[3mm]
{\large{\em{$^{a}${Department of Physics and Astrophysics, University of Delhi, Delhi, Delhi-110007, India.}}}}\\
{\large{\em{$^{b}${Department of Physics, Miranda House College , University of Delhi, Delhi, Delhi-110007, India.}}}}\\
{\large{\em{$^{c}${Department of Physics, Swami Shradhanand College, University of Delhi, Delhi, Delhi-110036, India.}}}}\\
{\large{\em{$^{d}${Department of Physics, Hansraj  College, University of Delhi,Delhi, Delhi-110007, India.}}}}\\
{\large{\em{$^{e}${Department of Physics, Pt.N.R.S.G.C.Rohtak, Maharshi Dayanand University, Rohtak-124001, Haryana, India.}}}}
\vskip 1pt
{\bf{Abstract}}: 
 We review on the theoretical status of intense laser orientation and alignment- a field at the interface between intense laser physics and chemical dynamics with the potential applications ranging from high harmonic generation, nano-scale processing and control of chemical reactions. The evolution of the rotational wave packet and its dynamics leading to orientation and alignment is the topic of the present discussion. The major part of the article basically presents  an overview on recent theoretical progress in controlling the orientation and alignment dynamics of a molecule by means of shaped laser pulses. The various theoretical approaches that lead to orientation and alignment ranging from static electrostatic field in combination with laser field(s), combination of orienting and aligning field, combination of aligning fields, combination of orienting fields, application of train of pulses {\it{etc.}} are discussed. It is focussed that the train of pulses proves to be quite efficient in increasing the orientation or alignment of a molecule without causing the molecule to ionize. The orientation and alignment both can occur in adiabatic and non-adiabatic conditions with the rotational period of the molecule into consideration. The discussion is mostly limited to non-adiabatic rotational excitation (NAREX) cases where the pulse duration is shorter than the rotational period of the molecule. A particular focus is put on the so called half-cycle pulse (HCP) and square pulse (SQP). The effect of ramped pulses is also under focus. The effect of collision on the various laser parameters is also given concern. We summarize the current discussion by presenting a consistent theoretical approach for describing the action of such pulses on movement of molecules. The impact of a particular pulse shapes on the post-pulse dynamics is calculated and analysed. In addition to this, the roles played by various laser parameters including the laser frequency, the pulse duration and the system temperature etc. are illustrated and discussed. The concept of alignment is extended from one-dimensional alignment to three-dimensional alignment with the proper choice of molecule and the polarised light. We conclude the article by discussing the potential applications of intense laser orientation and alignment.
\vskip2pt
\parindent 20pt
\tableofcontents
\section{Introduction}
 Light has powered us since the beginning of life on earth . Fundamental research based on properties of light has led many technical applications that have changed and shaped our social and personal life. Light, an electromagnetic radiation have wide range of wavelengths ranging from gamma rays (having wavelength less than $10^{-11}$to radio-waves (longest wavelength). Discovery of  light based  technologies like; discovery of fibre optics, lasers, telescopes, remote sensing satellites and many more, have brought revolution in field of communication, defence, health and  research in theses areas.

 With the advent of LASER (an acronym of 'Light Amplification by Stimulated Emission of Radiation') the pioneering research in high field lasers have made it possible to achieve high field strengths.  The $Nd^{3+}$ or the Nd:YAG laser, the first widely available intense laser has remained the workforce for many experiments with intensity range ($10^{13}-10^{14} W/cm^{2}$). With the invention of chirped pulse amplification(CPA) since 1985, it is now possible to generate intensities of the order of  1 PW (Petawatt) and beyond. Such intense lasers can produce the fields comparable or stronger than  inter-atomic fields.The atoms and molecules in such high  fields exhibit unusual properties which are important to understand physics of many   potential applications  like, studying  dynamics of ultra-fast phenomena, developing high frequency  lasers, analysing the properties of condensed matter and plasma under harsh conditions (of temperature and pressure).  Intense laser fields are also important to understand and control chemical reaction dynamics of atoms and molecules.  
  
 The interaction of laser with matter can be categorised into two groups {\it{viz.}} laser-induced processes and laser-assisted processes. The laser-induced processes are activated by a laser field and occurs for a threshold value of absorbed laser photons (in number) . The processes which belong to this group are  multi-photon and above-threshold ionization (MPI/ATI) of atoms, multi-photon and above threshold dissociation (MPD/ATD) of molecules, coulomb explosion, high-harmonic generation (HHG), orientation, alignment and coherent control of chemical and physical processes [\ref{Gavr}-\ref{OA4ex}]. The investigation of these processes revealed a number of phenomena that have practical application in different fields of research like holography, fibre optics, material science, telecommunication. The efficiency of the said processes depends on the atomic and molecular species considered and also on  various laser parameters like frequency, intensity, polarisation, pulse duration. On the other hand,in the laser-assisted processes laser  influence the species externally, however it may be strong depending on intensity of the laser field. Examples of Laser-assisted processes are; Compton scattering [\ref{Comp}], photoelectric effect [\ref{phot1},\ref{phot2}], electron diffraction [\ref{ED}] and collisions [\ref{vpras}-\ref{Yang2}]. \\

 The ionization and excitation of atoms by the charged projectiles like ions and electrons is a topic of great interest in atomic physics. The study of different aspects of electrons emission in collisions with ions at large impact energies, with collision velocity v higher than orbiting velocity $v_{0}$ of the electron in its initial bound state is quite popular both theoretically and experimentally [\ref{Br}-\ref{Ei}]. Study of ion-atom collisions in external electromagnetic field introduces new degrees of freedom which have strong influence on collision dynamics. These   collisions are satisfactorily explained through Floquet formalism. The Floquet approach is usually applied to the processes where field duration is large enough compared to laser-atom (molecule) interactions.

 Controlling the rotational dynamics of  atoms and molecules in laser fields is an exciting field and at low temperature  rotational transitions becomes more important than other degrees of freedom. These rotational transitions are primarily investigated under adiabatic and non-adiabatic conditions. The applicability of the condition depends on the duration of rotational period and the pulse duration. In  adiabatic rotational excitations(AREX), the laser pulse period  is larger than rotational period of interacting atom or molecule i.e. $T_{pulse}>\frac{5\hbar}{B}$, where $\hbar=\frac{h}{2\pi}$, with $h$ as the Planck's constant and B, is the rotational period of the molecule. Under such circumstances  interacting species behaves as it is in static  field  and  states created are stationary  states (pendular states) and the rotational excitation dynamics follows the pulse shape [\ref{Ca}]. In the non-adiabatic case {\it{i.e.}} non-adiabatic rotational excitation (NAREX), the pulse duration of laser  is much smaller than rotational period of molecule or atom i.e. $T_{pulse}<\frac{\hbar}{B}$) and interacting species  ends up in a rotational wave-packet.

Pulse duration of the laser is an important specification for  diagnosis of the system under consideration. Various methods to obtain wide range of pulse durations from nanoseconds, picoseconds, femtoseconds and even upto attoseconds  has been developed. Various outstanding reviews exist which brief about these methods [\ref{FK11r}-\ref{IA11r}].Development of such kind of  scientific research  on various aspects related to laser has opened new avenues  to explore the time evolution in a desired spectral regime leading to landmark discoveries in the subject of physics and chemistry.
 
The molecular alignments and orientations plays a significant role in many processes like  laser based isomerization[\ref{Dion}], scattering of molecules from surfaces [\ref{Tenner}], photodissociation by multiphoton processes [\ref{Bandrauk1}], laser focussing techniques for  nano scale design [\ref{TSeideman}, \ref{Stapefeldt1},\ref{Stapefeldt}]]. Experimental detection of impulsive alignment of  linear molecule ($I_2$) (for first time ) was reported in year 2001 and 2002 [\ref{Rosca} ]. Single pulse nonadiabatic alignmrnt for asymmetric molecule (Iodobenzene ) has been studied theoretically and experimentally [\ref{EP2}], which is mainly centred  around observing maximum alignment, however they have also observed rotational revivals  after the pulse is turned off.Initial experimental studies related to laser induced alignment are done by Nirmand {\it{et al.}} [\ref{DNL}]  and Kim {it{et al.}} [\ref{PMF}]. 
 
To enhance field free alignment (non adiabatic alignment or  short pulse alignment or impulsive alignment ) two cycle laser pulse is more advantageous than single pulse. First experimental verification of this fact was established in 2004 by Bisgaard {\it{etal.}}  {\ref{Bisgaard}} for asymmetric molecule Iodobenzene ($C_{6}H_{5}$). They have found that two pulse alignment is enhanced when second pulse is made three times stronger than former pulse and attuned near the time when alignment due to previous pulse is maximum. Renard etal [\ref{Renard}] have reported  experimentally 'nonintrusive' after pulse alignment using short laser pulse polarisation technique. They have also studied planar delocalisation for $CO_{2}$ molecule. 

Alignment is a measure of order in which molecules are arranged or placed with respect to axis  fixed in space, however orientation is related  to specification of direction with respect to the space fixed axis. Techniques for  attaining orientation or aligninment are  mainly based on collisions, application of laser and optical fields and static fileds or combination of static and laser fields(hybrid fields)[\ref{Stapefeldt}].

There are many early interesting studies done both experimentally and theoretically to orient the molecule. Static electric field  based studies for orienting a molecule include hexapole state selection [\ref{Kramer},\ref{Hain}], brute-force technique[\ref{Loesch},\ref{BFDR}], which are applicable only for polar molecules. They are beneficial because they arrange molecules in head-tail fashion. These studies was done for symmetric and asymmetric polar molecules. However they have one limitation that these approaches can not be applied for non polar molecules. The optimal control scheme and coherence  control  schemes are also  effective technique to deal with orientation of molecules [\ref{RSJUDSON},\ref{JJVrakking}]. Simultaneous Interaction of polarisability and permanent dipole moment with two color infrared(IR) laser, leads to symmetry breaking in molecules and leads to enhanced orientation in a triatomic system (HCN)[\ref{CMDION}].
For nearly two decades, theoretical and experimental studies on orientation and alignment of molecules , to prepare focused   molecular beam with high anisotropic distribution of orientations are cusp of interesting studies. [{\ref{Bern}-\ref{JB3}] references discusses useful techniques to orient and align the molecules.In these studies   static electric or nonresonant laser fields ha been utilised to orient or align the molecule.At low temperatures inhomogeneous DC field deflects the  molecules most which resides in ground states or even can isolate them. Such deflected or isolated molecule can be used as targets for experiments. Strongly  defected molecules produce higher degree of alignment and orientation than  less deflected molecules[\ref{SVH}]. It is possible to achieve  field-free molecular orientation in combination of DC electric field and laser pulse, when laser field  is adiabatically turned on and rapidly non-adiabatically  turned off [\ref{AGob}] .
  
\indent Several applications are based on rapid orientation of the molecule and the AREX may not be  prove to be suitable choice. In NAREX, the application of a short Half-Cycle Pulse (HCP) can lead to fast orientation of polar molecule [\ref{Dion1},\ref{Machholm}]. The advantage of linearly polarised HCP is that it performs a short oscillation half-cycle followed by a long but much weaker tail of opposite polarity. Such a short duration pulse transfers momentum $\Delta p$ (a kick) to the system [\ref{TSNE}] and the transferred momentum is proportional to the pulse strength and its duration. The orientation thus obtained after the pulse is field-free and can be desirable in some situations. The resulting molecular orientation, doesn't sustain for long as the effective duration of the HCP is much smaller than the rotational period of the molecule. Hence, once the pulse has passed by the molecule, it evolves in a field-free manner, oscillating from oriented to non-oriented configurations. Thus, the time average of the orientation over the rotational period of molecule vanishes. The post pulse orientation of the molecule is desirable in some situations but depends on limitations imposed by the time resolution of the phenomenon under study or on the experimental capabilities. So, it is desirable to have some method that is capable of inducing strong and sustainable molecular orientation in non-adiabatic regime.The combination of a circularly and linearly polarized laser pulses to control field-free molecular orientation has been reported by Chi {\it{et al.}} [\ref{Chh}]. The application of circularly polarised laser pulses aided by linearly polarised resonant pulse, in resonance with molecular rotational transitions, prepares a rotational wave-packet thus populating few lowest rotational eigenstates with magnetic quantum number $M=0$. The application of a particular series of linearly polarized pulses may lead to enhancement of orientation. Zhao {\it {et al.}} [\ref{ZY}] have studied  the effect of field free orientation   on pre-dissociation  of NaI molecule. 

When laser pulse duration is much smaller than molecule's rotational period alignment is non adiabatic in nature [[\ref{JO},\ref{NEH}]], that is molecule is still aligned when laser pulse is switched off.Due to this reason non-adibatic alignment is also given the name of  field-free alignment [\ref{EP}]. First experimental verification for non-adiabatic alignment was done by Vrakking {\it{et. al.}} \ref{Rosca}]. 

When  strong aligning field interacts with a molecule, it is important to understand that  how rotational excitation dynamics and  the aligned pendular states' behaviour evolve with different field parameters. Such kind of understanding  is important for many applications like trapping the molecules in pendular states and other applications realted to this field [\ref{BAZ}-\ref{BFri}]. Photodissociation has also been exhausted to attain alignment [\ref{HSak}].

Linearly polarized laser fields only cause one dimensional (1D) alignment in linear and symmetric top molecules. On the other hand, a asymmetric-top molecule when exposed to elliptically polarised light leads to three dimensional(3D) adiabatic alignment [\ref{JJLK}] and it is confined in all three dimensions in space fixed axis and the molecule is not left with any option of free rotation.Improved degree of  molecular orientations has been obtained in literature by various approaches like pulse train [\ref{Wu}] ,by maximizing optimal fields [\ref{Salomom}], by slow turn on and quick turn off of the two color laser field [\ref{Muramatsu}].For aligning CO molecule, super Gaussian pulse has been found to be  more favourable than standard Gaussian pulse  [\ref{QCH}]. However super Gaussian pulse does not favour the adiabatic orientation at pulse duration as   does the standard Gaussian pulse [\ref{JSLIU}]. FCN molecule has been found to posses high molecular alignment under THz half-cycle pulse than that by a Thz few-cycle pulses            [\ref{QJLIU}].
  
 Phase or amplitude modulation techniques are exploited to generate multiple train pulses which are efficient in enhancing population inversion in molecular systems. These sequences of pulses are also beneficial for multi-photon pumping  [\ref{Warren1}]. Spectral properties of multilevel atoms or molecules using  train of ultrashort laser pulses of different shapes  are explored theoretically [\ref{Felinto}]. A pulse train can be of single pulse  with attosecond duration [\ref{Mauritsson}] or multiple cycle pulses [\ref{Rausch}]. Cryan {\it{et al.}} [\ref{James}] have observed  experimentally that linearly polarized multiple pulses i.e. a train of eight pulses (in which each pulse has equal energy) enhances  alignment of  nitrogen at Standard Temperature and Pressure(STP) due to Raman excitations which are impulsive in nature. It has been observed by Gogyan {\it{et al.}} [\ref{Gogyan}]  that ultrashort laser pulse train combined with weak coupling  control fields produces the superposition of the atomic or molecular states in the similar way as ultrafast laser pulse train does.Using this mechanism destructive effects (which occurr in strong field ), like ionization can be avoided.

When polar or non polar linear molecules are placed in combined action of static electric and linearly polarised laser field degree of orientation gets enhanced, theoretical explanation of this observation for first time was given  in references [\ref{B1},\ref{B2}]. They have discussed dynamics of the molecules when both the fields are collinear or non-collinear fields. Later on theses studies were extended for symmetric [\ref{HF1}] and asymmetric top molecules [\ref{O1}]. 
Observations of alignment and orientation in mixed fields (laser and static electric field) in benzonitrile (asymmetric top) molecule claimed that  when both the fields are perpendicular the alignment is purely adiabatic in nature, however for tilted fields pure adiabatic description is not correct, hence one need to give non-adiabatic description of the field dressed  molecule to understand the complete dynamics [\ref{O2}]. An ensemble of aligned molecules in presence of IR and UV laser pulse can be exploited to get a sub-ensemble of oriented molecules [\ref{A1}]. Other studies which throw a light to get enhanced orientation are presented in references [\ref{B3}-\ref{S1}].  Three dimensional alignment of 3,4-dibromothiophene has been confirmed experimentally in combined elliptically polarised light and static electric  field [\ref{H2}]. Liu {\it{et al.}} [\ref{JLQ}] have studied orientation and molecular alignment driven by elliptically laser pulses. They have obtained 3D  field free molecular alignment, while field free molecular orientation is realized only along two directions.  

Other studied such as orientation of organoxenon (H-Xe-CCH) molecule in the combined  laser and static electric fields [\ref{V1}],  field-free orientation of  molecules (symmetric-top and other) by tetrahertz(THz) laser pulses [\ref{PBK}-\ref{PBK2}]  at high temperature are also paid attention.Field free alignment and rotational dynamics of linear molecules has been investigated  by strong-field ionization  [\ref{GKN}]. Phase shaped femtosecond laser pulses has been used to study selective excitaions in rotational dynamics and molecular  orientation {\it{et al.}} [\ref{YHS}]. Lemeshko etal have presented well the literature related to manipulation of molecules in various fields [\ref{MLV}]. Koch {\it{et al}} have thrown light on work done till now on  theoretical and experimental aspects of quantum control of molecular rotations in short laser pulses [\ref{QCR}]. Lin etal [\ref{KL}] have presented recent progress in the field of molecule's unidirectional rotation ,rotational echoes. They have also discussed orientation of asymmetric top and chiral molecules in three dimensions.  

In section 2, we have given theoretical description on the terms NAREX, orientation and alignment. The section discusses in detail the conditions under which NAREX occurs and how NAREX leads to orientation and alignment under field-free conditions which proves to be quite useful for various practical applications. Theoretical methods like direct method, Magnus Approximation, (t,t') method and Split operator method, for solution of time dependent Schr\"{o}dinger equation (TDSE) are also discussed. The importance of pulse shape is also under concern. 

 In section 3, we discuss various approaches that control the molecular rotation. There exist various techniques which can be used to control molecular orientation and alignment. The orientation can be obtained by static electric field in combination with the delayed pulses and also by ramped pulses. The suitable combination of orienting and the aligning pulse (or pulse train) can also control molecular orientation and alignment. The concept of 2-D alignment is also one of the useful technique to control molecular alignment. The near-adiabatic orientation and alignment by mixed-field also proves fruitful in handling orientation and alignment. 
 
In section 4, we focus on one of the most commonly studied laser-assisted process, i.e. the collision process, and how it modifies the rotational excitation of a diatomic molecule.

In section 5, few applications of orientation and alignment are discussed.

In section 6, we present the outlook and conclusion of the detailed work.

\section{Theoretical Consideration}

\subsection{Theoretical Description of NAREX, Orientation and Alignment}

The  non-adiabatic rotational excitation (NAREX) dynamics is of fundamental importance in various natural sciences, hence a lot of experimental and theoretical studies have been done in order to explore it in various systems of interest [\ref{Markevitch12}-\ref{Baek}].
In processes like nuclear rearrangement [\ref{Kosmidis}], fragmentation to nuetral products [\ref{Dewitt}],internal conversion [\ref{Anderson},non-resonant electronic excitation [\ref{Boyle}] NAREX plays important role.
 
Orientation and alignment are two important parameters for manipulating  spatial arrangements of molecules. These observables  establish a connection between laboratory frame and molecular frames to understand various molecular properties.

Alignment is actually, confining the molecular axis with reference to space fixed axis. However orientation is related to breaking the inversion symmetry. In case of Alignment there is no preference for direction and it appears as double headed arrow. On the other hand, in orientation emphasis is on direction and it behaves as a single headed arrow. Figure  (\ref{oa}) illustrates orientation and alignment in case of a diatomic molecule.  Quantitative measures of alignment and orientation parameters are $\langle cos^{2} \theta \rangle$  and $\langle cos\theta \rangle$, respectively.  In case of perfect alignment  $\langle cos^{2} \theta \rangle$  is equal to 1, for isotropic distribution of molecules it has values of 0.33. However in case of complete anti-alignment  $\langle cos^{2} \theta \rangle$ value is 0, and all molecules are perpendicular to the alignment axis. Orientation parameter $<cos \theta> $ has a value of +1 or -1 in case of perfect orientation  or perfect anti-orientation, respectively. If $<cos \theta>$ value  is 0, the molecules are not oriented [\ref{Kev}]. 

\begin{figure}
\vskip 7.4cm
\hskip 1.2cm
\includegraphics{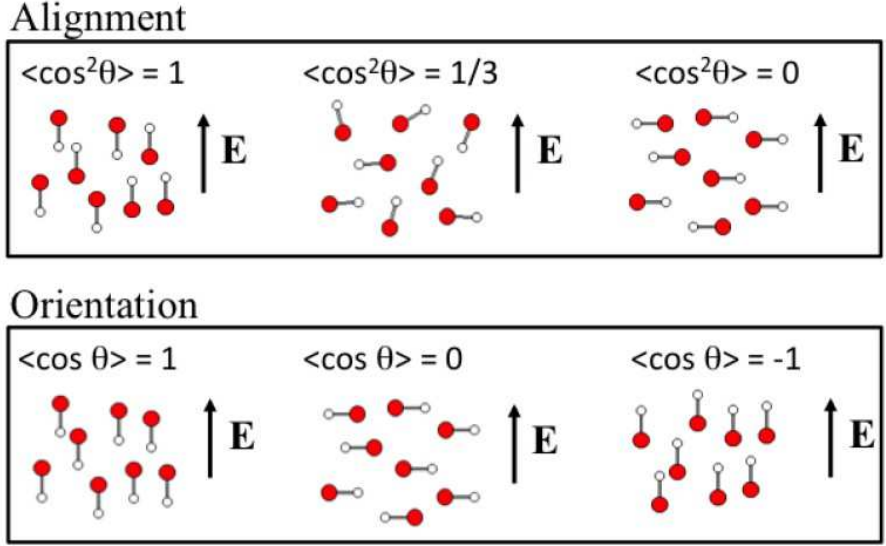}
\caption{Illustration of a sample of diatomic molecules placed in an electric field E. In the upper panel the molecules are: aligned along the field (left), isotropic (middle), anti-aligned, $i.e.$ their molecular axes are perpendicular to the electric field (right). In the lower panel the molecules are: oriented along the field (left), isotropic (middle), oriented
in the opposite direction with respect to the electric field (right). The values of $\langle cos^{2}\theta \rangle $ and $ \langle cos \theta \rangle $ signify the degree of alignment and orientation, respectively. The angle $\theta$ corresponds to the angle between the electric field and the molecular axis. }
\label{oa}
\end{figure}

\subsubsection{Evolution of Rotational Wavepacket}

Molecular rotation caused by laser pulse produces  rotational wavepackets  which re-phases or dephases depending upon  fundamental period $T_{rev}=\frac{1}{2Bc}$. At revivals application of additional laser pulse  changes the dynamics of these wave packets significantly. In  a freely aligned wavepacket  periodic revivals occur  after every fundamental period  $T_{rev}$. [\ref{Doo}].\\  For first time rotational revivals were observed for gas cells [\ref{Heri}].

Interaction of a non-spherical polar molcule with $sin^{2}$  laser pulse (with electric field polarization along z-axis) is given by TDSE as :

\begin{equation}
\label{li1}
i\frac{d\psi}{dt}(\theta,\phi,t)=[B\hat{J^{2}}-U_{0}(t)cos^{2}\theta]\psi(\theta,\phi,t),
\end{equation}

where, $\theta$ represents  the angle between the electric field polarization and the molecular axis, $B\hat{J^{2}}$ is free rotor Hamiltonian  ( \lq B\rq is rotational constant  and  $\hat{J^{2}}$ is the squared angular momentum operator). The term, $U_{0}(t)cos^{2}\theta$ is the interaction term with $U_{0}(t)=\frac{1}{4}\Delta\alpha E_{0}^{2}  sin^{2}(\frac{\pi t}{\tau}$, where, $\tau$ is pulse width.

The evolution of rotational wavefunction can be obtained by expanding time dependent wave function in $|J,M>$ basis as :

\begin{equation}
\mid \psi(t)\rangle = \sum _{J,M} A_{J,M}(t) \mid J,M \rangle.
\label{nw}
\end{equation}

In above expression, $A_{J,M}(t)$ is expansion coefficient and $\mid J,M \rangle$ is free rotor wavefunction.

Using equation (\ref{nw}), time dependent Hamiltonian, $H(t)=[B\hat{J^{2}}-U_{0}(t)cos^{2}\theta]$  takes the form : 

\begin{eqnarray}
\label{li2}
\langle J,M \mid H(t) \mid \psi(t) \rangle & = & BJ(J+1)A_{J,M} - U_{0}(t)C_{J,J+2,M}A_{J+2,M} \nonumber \\
& & 
-U_{0}(t)C_{J,J,M}A_{J,M}-U_{0}(t)C_{J,J-2,M}A_{J-2,M},
\end{eqnarray}

with 

\begin{eqnarray}
C_{J,J,M}&=&\langle J,M \mid cos^{2}\theta \mid J,M \rangle , \\
C_{J,J+2,M}&=&\langle J,M \mid cos^{2}\theta \mid J+2,M \rangle ,\\
C_{J,J-2,M}&=&\langle J,M \mid cos^{2}\theta \mid J-2,M \rangle.
\end{eqnarray}

In expression (\ref{li2}), for time evolved Hamiltonian odd and even J states do not couple due to symmetry of $cos^2 \theta$, angular potential term. Also, M states coupling is forbidden due to cylindrical symmetry present in angular potential term. 

In case of field-free evolution of the wavepacket, the time-dependent wave function becomes 
\begin{equation}
\psi(t)=\sum_{J}A_{J,M}e^{-iE_{J}t} \mid J,M \rangle,
\end{equation}
where, $E_{J}$ is the eigenenergy and $E_{J}=BJ(J+1)$.\\

\subsubsection{Measure of Alignment}

\indent A standard way to define the degree of alignment of rotational wave packet is defined by average value of $cos^{2}\theta$,

\begin{equation}
\langle cos^{2}\theta \rangle=\langle \psi \mid cos^{2}\theta \mid \psi \rangle.
\end{equation}

The value $\langle cos^{2}\theta \rangle=1$, implies perfectly peaked angular distribution along the poles $\theta=0$ and $\theta=\pi$; $\langle cos^{2}\theta \rangle=0$ implies distribution peak along the equator where $\theta =\pi/2$ and $\langle cos^{2}\theta \rangle = \frac{1}{3} $, shows an isotropic distribution evenly distributed across all $\theta$.\\

\indent In case of field-free propagation, the time-dependent measure of alignment is given by:

\begin{eqnarray}
\langle cos^{2}\theta \rangle (t)&=& \langle \psi(t) \mid cos^{2}\theta \mid \psi(t) \rangle \\
				& =& \sum_{J} [{\mid A_{J,M} \mid}^{2} C_{J,J,M} \nonumber\\
& & + \mid A_{J,M} \mid \mid A_{J+2,M} \mid cos(\omega_{J} t+\phi_{J,J+2})C_{J,J+2,M} ],
\end{eqnarray}

where,

\begin{equation}
\omega_{J}=E_{J+2}-E_{J},
\end{equation}
 
and $\phi_{J,J+2}$ represents the relative phase between the states $\mid J,M \rangle$ and $\mid J+2,M \rangle$ at the initial stage of the field-free evolution.

\indent For a system of molecular gases interacting with a linearly polarized infra-red laser pulses, the laser pulse interacts with the molecular gas and the system is assumed to be a thermal ensemble  which is characterized by a temperature $T$. In quantum terms, such a system is explained as a statistical mixture of angular momentum states. Here, the molecule finds itself in a definite state of angular momentum $\mid J,M \rangle$, where $J$ is the orbital angular momentum with $J=0,1,2,3,.........$ and $M$ is the projection of the angular momentum onto the z-axis of the coordinate system with $M=-J,-(J-1),.............(J-1),J$. The coordinate wave-function of such basis are the spherical harmonics $Y_{M}^{J}(\theta,\phi)$. The distribution of the angular momentum amongst various molecules in the gas is represented by the Boltzmann distribution.\\

\begin{equation}
P_{J} \sim (2J+1)e^{-E_{J}/k_{B}T},
\end{equation}

with $E_{J}$, as rotational energy of the state $\mid J,M \rangle $. Here, $(2J+1)$ represents the degeneracies within a given $J$ level because of different $M$ sub-levels.\\

\indent	For the case of homo-nuclear diatomics, an additional factor $g_{J}$ in the Boltzmann distribution arises from nuclear spin statistics [\ref{Her}]. 

\begin{equation}
P_{J} \sim g_{J}(2J+1)e^{-E_{J}/k_{B}T}.
\end{equation}

This factor is the measure of relative weight between even and odd states. For such cases, the measure of alignment is given by:

\begin{equation}
\langle cos^{2} \theta \rangle (t)=\frac{\sum_{J,M} g_{J}(2J+1)e^{-E_{J}/k_{B}T}\langle cos^{2}\theta \rangle(t)_{J,M}}{\sum_{J}P_{J}}.
\end{equation}

\subsubsection{Adiabatic and Non-adiabatic Alignment}

If the laser pulse duration is much larger than  the molecule's  rotational period adibatic alignment occurs in molecules. Initial studies to achieve the alignment in molecular systems are adiabatic [\ref{HSA}-\ref{JJL1}].  
Molecules in  strong laser fields gets polarized and its  highest anisotropic  polarizability axis is in  the direction of laser field polarisation [\ref{NJJ}]. When  linear molecules are placed in intense laser fields they always have anisotropic  polarizability and the maximum polarizability is directed  along the molecular axis. Due to adiabatic interaction of the molecule with laser field pendular states are created. In molecular dynamics laser field  polarisation plays an important role and it acts as tool to govern the desired phenomena [\ref{ADBan},\ref{CMDion},\ref{HJLoesch}]. As the laser field is turned off, the molecular system returns  to its initial state [\ref{BFri}]. The adiabatic alignment can be attained by nanaosecond laser pulses (up to 100 ns). Larsen {\it{et al.}} [\ref{Larss}] and Sakai {\it{et al.}}  [\ref{Saka}] have investigated adiabatic alignment dynamics in nanosecond laser filed. Adiabatic strategy usually produce high degree of  alignment in presence of laser field, however presence of this strong  laser field may distort processes  under investigation.

When laser pulse duration is much shorter than the rotational period of a molecule, molecules align non-adiabatically [\ref{BAZ},\ref{JO}]. In this impulsive alignment, short laser pulse produces superposition of rotational states which is coherent in nature because of transfer of large angular momentum to the molecule after interaction. This superposition of rotational states is called as rotational wave packet which re-phases and de-phases when  laser field is switched off [\ref{Averb}-\ref{GHLEE1}]. It is purely quantum mechanical effect. This rotational wave packet is responsible for the rotation of the molecule even in the absence of  laser field.
Non-adiabatic alignment theory in detail  is  well articulated  in  [\ref{Se},\ref{NEH}]. Rosca-Purna and Vrakking [\ref{Rosca}]  demonstrated  experimentally  nonadiabatic alignment in molecules in 2001  for first time. 
  
 Non-adiabatic alignment is short lived, however occurs in field free conditions  in the absence of the impulsive interaction, is important  for further use in  many applications [\ref{Se}]. In reference [{\ref{KoMi}] non-adiabatic alignment of asymmetric top molecule is investigated using optical centrifuge. Non-adiabatic or impulsive alignment is usually achieved by femtosecond and picosecond laser pulses.In impulsive fields the rotational dynamics of the molecule depends on characteristic rotational period of the molecule.
 
A hybrid  technique based on  adiabatic and nonadiabatic strategies has been  developed by Yan
et al. [\ref{ZCY}]. In this strategy the laser field is switched on slowly  and the switched off quickly. Here pendular states are created adiabatically, which rephase and dephase as in nonadiabatic case under field free conditions. Maximum alignment obtained during each revival is same as it occurs in pure adibatic conditions.  
Torres {\it{et al}}. [\ref{Torres}] have  detailed the quantum mechanical theoretical description of adiabatic and non-adiabatic  alignment  to compare and analyze the alignment dynamics.They have also investigated the impact of temperature, laser intensity and pulse duration in order to realize the  conditions for optimum alignment.   
  
All molecules do not have permanent dipole moment, however all molecules posses reasonable polarisation in suitable field strengths. Hence all molecules can be aligned easily in comparison to orientation because of the presence of second order effect due to polarisability. Molecules will be aligned in one dimension [\ref{MLI},\ref{Luo}], two dimension  or in three dimensions [\ref{Makh}-\ref{Rouz}]   is   decided by type of polarisability and symmetry of the aligning field. Linearly polarised laser fields create 1-D alignments [\ref{MLI},\ref{Luo},\ref{ARS}]. Daem etal [\ref{Daem}] has realised two dimensional alignment using Short elliptically polarised laser pulses. Three dimensional alignment leads to confinement of molecular axis along space fixed axis. Experimental studies which investigated three dimensional alignment are given in references [\ref{Under},\ref{JJLK},\ref{Lee}]. Theoretical nonadiabatic description of three dimensional alignment is given in reference [\ref{Take}].

\subsection{Theoretical Methods}

In order to understand basic physics behind molecular systems the time dependent  Schrodinger equation is one of the fundamental equations. Hence in  modelling and simulations of molecular dynamics, its solutions are of central importance and hence many numerical techniques has been developed by scientific community to  solve it numerically or analytically (wherever it is possible) .

 \subsubsection{Direct Solution of TDSE (Time-dependent Schr\"odinger Equation)}
 
For intensities less than $10^{13}W/cm^{2}$, the perturbation theory works well. The non-perturbative behaviour has been noticed for intensities above $10^{13}W/cm^{2}$. The various non-perturbative methods like essential states expansion, R-matrix, Quasi energy technique {\it{etc.}} are based on the assumption that the Hamiltonian of the atomic system in the laser field is periodic in time which might not be possible for a realistic case.

However with these non-perturbative methods it is difficult to solve analytically Time dependent Schr\"odinger  equation for most of the systems. Hence by making required approximations to include temporal, spatial and propagation effects, these methods  give reliable results. However every method has its own limitations. An alternative to these non-perturbative methods is the direct integration of the TDSE in one dimension and three dimension. Direct integration is possible in all regimes of laser intensity and frequency. Also, It is valid for most of pulse shapes and hence various pulse shape effects can be explored. However in most of integration algorithms spatial grid size and time step size are two important parameters to decide their worth, to attain accurate results it is preferred to use large spatial grid  having small spatial separation and large number of small step size.

One dimensional integration technique for first time was practised in late seventies [\ref{Gel}-\ref{Aus}]. Later on  Elberly {\it{et al.}} [\ref{Ebe}] and others [\ref{Ree1}-\ref{Pro}] also developed more algorithms to deal with one dimensional  Schr\"odinger   equation. Main disadvantages of these methods is the exaggeration of singularity in Coulomb potential. These algorithms can not be applied to a system interacting with magnetic field or circularly polarised field and are valid only for linearly polarised light. However these methods are computationally very fast and hence as a result large number of parameters can be evaluated by using them. 

One dimensional direct integration methods are an approximation of three dimensional model. Rae {\it{et al.}} [\ref{Rae}] have shown that due to exaggerate picture of atomic core, one dimensional models overestimate the ionization rate. Hence in order to get complete understanding of dynamics of the system three dimensional algorithms should be implemented. Three dimensional models used to study atom laser interactions have been developed in early nineties [\ref{Kul1},\ref{Kul2}]. Other studies on time evolution of one electron atom and photo-ionization  are presented well in literature [\ref{LaG},\ref{Ros},\ref{Kra}]. In these studies the time dependent wave function is expanded in terms of angular momentum components which are coupled to the laser fields. Recently Wells {\it{et al}} [\ref{Wells}] have developed a computationally fast integration method to study time evolution of electronic wave function. They claim  that it is possible to apply higher-order approximations to spatial as well as  temporal derivatives than other more existing  rigid methods . Three dimensional methods in general computationally more expensive than one dimensional methods but still they provide comparatively deep insight into time evolved   dynamics of the atoms or molecules in  external fields. An alternative to reduce the size of calculation is to make use of linearly polarized light such that the system has axial symmetry in the dipole approximation and dimension reduced to one [\ref{Per},\ref{Kul2}].

\subsubsection{Magnus Approximation}

Time dependent Schr\"{o}dinger equation for a time evolution operator is:
\begin{equation}
\iota \hbar \frac{\partial U(t,t_{0})}{\partial t}=[H_{0}+V(t)]U(t,t_{0}).
\label{er2}
\end{equation}

In Schr\"{o}dinger's picture, an isolated system is described by time displacement operator: 

\begin{equation}
U(t,t_{0})= exp(- \iota(t-t_{0} H/\hbar)).
\label{en}
\end{equation}

In presence of external fields, it is customary to represent $U(t,t_{0})$ time displacement operator in perturbation series as follows:

\begin{equation}
U(t,t_{0})= 1- \frac{\iota}{\hbar} \int_{t_{0}}^{t} dt_{1}H(t_{1})U(t_{1},t_{0}) \nonumber
=1-\frac{\iota}{\hbar} \int_{t_{0}}^{t}dt_{1}H(t_{1})+ (\frac{\iota}{\hbar})^{2}\int_{t_{0}}^{t}dt_{1} \int_{t_{0}}^{t_{1}}dt_{2}H(t_{1})H(t_{2}).
\label{en1}
\end{equation}

Above expansion is obtained by iterating equation ({\ref{en}}). The approximation in equation (\ref{en1}), on truncation at any point is not unitary and is valid for small $(t-t_{0})$. However  defining an anti-Hermitian operator as: 

\begin{equation}
exp[A(t,t_{0})]= U(t,t_0).
\label{en2}
\end{equation}

Any approximation, which conserves anti-Hermitian character of the operator $'A'$ makes $U(t,t_{0})$ an unitary approximation. 

According to Magnus, $'A'$ is an infinite series with 'nth'  term  as  sum of integrals of n-fold
multiple commutators of time dependent $H(t')$ [\ref{Magnus}]. Exponential formulation of time displacement operator has been   developed independently by Robinson  [\ref{Robinson}], because when he derived it, he was unaware of Magnus work. However Pechukas {\it{etal}} [\ref{Pech}] realised the importance of Magnus work and hence, derived  Magnus expansion and discussed its convergence in detail for time evolution operator. For simple problems Magnus approximation reduce the efforts to obtain exact solutions,Pechukas {\it{etal}} have demonstrated it by doing calculations for harmonic oscillator. Importance of Magnus formulation for molecular-scattering problem is also discussed in [\ref{Pech}].

Review by Blanes etal [\ref{Blanes}] discusses all developments in literature related to  Magnus approximation till 2008. They have also detailed its importance and applications in various fields of physics. Major contributions of Magnus approximation  which improved and widened the scope of studies and understanding  are coulomb excitations in strong fields [\ref{Robinson}], heavy-ion impact [\ref{Eicher}], inner shell excitations in atom-ion collision induced by rotational dynamics [\ref{Wille},\ref{Wille1}], pressure broadening phenomena in rotational spectra[\ref{Cady}], multiphoton excitations in sparse systems where rotating wave approximation fails [\ref{Sheck}], electron-atom collisions in strong fields [\ref{Hyman}]. In Schrodinger picture, Magnus expansion face convergence problem [\ref{Salzman}], however in interaction representation,  Magnus expansion has been found to converge for two level system [\ref{Maricq}].

Henriksen {\it{et al}} [\ref{NEH}] have derived analytically Magnus approximation based  propagator, for studying  impulsive excitation dynamics of the linear molecule in an electromagnetic field.

Electric dipole moment $\mu$  and field  coupling is given by: 

\begin{equation}
V(t)= -\mu(R)E_{0}a(t)cos(\omega t),
\label{er1}
\end{equation}

with, a(t) as envelope function of the pulse centred at $t=t_p$. The time dependent rotational  dynamics of the system using equation ($\ref{er2}$) is derived using the pulse propagator $U_p$  [\ref{Smith}].

In order to solve time dependent Schr\"{o}dinger equation (($\ref{er2}$), pulse propagator in interaction picture is defined as: 

\begin{equation}
U(t_{f},t_{i})=U_{0}(t_{f},t_{p})U_{p}(t_{f},t_{p},t_{i})U_{0}(t_{p},t_{i}),
\label{er3}
\end{equation} 

which means,

\begin{equation}
U_{p}(t_{f},t_{p},t_{i})=U^{\dagger}_{0}(t_{f},t_{p})U(t_{f},t_{i})U^{\dagger}_{0}(t_{p},t_{i}).
\label{er4}
\end{equation}

Here, $U_{0}(t,t_{p})$ is the time evolution operator for the free molecule, thus:

\begin{equation}
\iota \hbar \frac{\partial {U^{\dagger}_{0}(t,t_{p})}}{\partial t}=-U^{\dagger}_{0}(t,t_p)H_{0} .
\label{er5}
\end{equation}

Using eqns.(\ref{er2}) and (\ref{er5}), we obtain:

\begin{eqnarray}
\iota \hbar \frac{\partial U_{p}}{\partial t_{f}} & = & U_{0}^{\dagger}(t_{f},t_{p})V(t_{f})U(t_{f},t_{i})U_{0}^{\dagger}(t_{p},t_{i}) \nonumber \\
&=& U_{0}^{\dagger}(t_{f},t_{p})V(t_{f})U_{0}(t_{f},t_{p})U_{p}(t_{f},t_{p},t_{i}).
\label{er6}
\end{eqnarray}

As $U(t_{f},t_{i})U_{0}^{\dagger}(t_{p},t_{i})=U_{0}(t_{f},t_{p})U_{p}(t_{f},t_{p},t_{i})$,  according to eqn.(\ref{er4}). Hence we have:

\begin{equation}
U_{p}(t_{f},t_{p},t_{i})=ex\hat{p}\Bigg[-\frac{i}{\hbar} \int_{t_{i}-t_{p}}^{t_{f}-t_{p}}e^{\frac{iH_{0}\tau}{\hbar}}V(\tau+t_{p})e^{\frac{-iH_{0}\tau}{\hbar}} d\tau \Bigg],
\label{er7}
\end{equation} 

where, $ \hat{} $ refers to the time ordering of the integrals. Above  expansion of the propagator is independent of the form of $V(t)$. At $t_{p}=0$, $U_{p}$ represents  the propagator in the  standard interaction picture  [\ref{Merz}].

On neglecting time ordering in eqn.(\ref{er7}), it takes the form:
 
\begin{equation}
U_{p}(t_{f},t_{p},t_{i})=exp\Bigg[-\frac{i}{\hbar} \int_{t_{i}-t_{p}}^{t_{f}-t_{p}} e^{\frac{iH_{0}\tau}{\hbar}}V(\tau+t_{p})e^{\frac{-iH_{0}\tau}{\hbar}} d\tau \Bigg].
\label{er8}
\end{equation}

The integrand $U_{p}$ can be evaluated by the relation [\ref{Merz}]:

\begin{equation}
e^{-A}Be^{A}=B+[B,A]+\frac{1}{2!}[(B,A),A]+..............,
\label{er9}
\end{equation}

Using eqn.(\ref{er9}), the Eq.(\ref{er8})takes the form:

\begin{equation}
U_{p}(t_{f},t_{p},t_{i}) \\
=exp\Bigg[\frac{\iota}{\hbar}\mu(R).E_{0}F_{0}(\omega) \\
 -\frac{\iota}{\hbar^2}[\mu(R).E_{0}.\iota H_{0}]F_{1}(\omega) \\
-\frac{1}{2!} \frac{\iota}{\hbar^3}([\mu{R}.E_{0},H_{0}],H_{0}) F_{2}(\omega) \\
+ .......\Bigg]
\label{er10}
\end{equation}

and commutators are Hermitian. 

The term $F_{n}$ has the form:

\begin{equation}
F_{n}(\omega)=\int_{-\infty}^{\infty}\tau^{n}a(\tau + t_{p})cos[\omega(\tau+t_{p})]d\tau,
\label{er11}
\end{equation}

where $n=0,1,2, ....$. The pulse is peaked at $t_{p}$ and by considering final times $t_{f}$  larger than the pulse duration, integration limits in Eq.(\ref{er11}) could be extended to infinity.

\indent The orientation and alignment parameter can be obtained using the well known relation [\ref{Merz}]:

\begin{eqnarray}
e^{\iota(\frac{A}{\hbar})cos\theta}&=&\sum_{l=0}^{\infty}(2l+1) i^{l}j_{l}(\frac{A}{\hbar})P_{l}(cos\theta) \\ \nonumber
&=&\sum_{l=0}^{\infty}{\sqrt{4\pi(2l+1)}}i^{l}j_{l}(\frac{A}{\hbar})Y_{l,0}(\theta,\phi),
\end{eqnarray}

with, $A=\mu(R_{e})E_{0}F_{0}(\omega)$, $P_{l}(cos\theta)$ are Legendre polynomials and $j_{l}(\frac{A}{\hbar})$ are spherical polar Bessel's functions [\ref{NEH}]. Hence with help of pulse propagator $U_p$, finally excitation dynamics, orientation and alignment dynamics can be  discussed in detail [\ref{NEH}].

\subsubsection{ $(t,t')$ Method}

In $(t,t')$ method [\ref{UPN}], computational techniques developed for independent Hamiltonians are used to investigate the time dependent Hamiltonians. This method is based on the idea of using generalised Hilbert space [\ref{Sambe}, \ref{Howland}] for solving time dependent Schr\"{o}dinger equation as time independent equation.This extended  Hilbert space introduces time as an additional coordinate. Hence in integration of time dependent Schr\"{o}dinger equation time ordering operator is not required.

\indent TDSE solution offered by $(t,t^{\prime})$ method for an initial state $\Psi(x,0)$ is of the form [\ref{UP}]:

\begin{equation}
\Psi(x,0)=\int_{-\infty}^{\infty}dt^{\prime}\delta(t^{\prime}-t)\Phi(x,t^{\prime},t),
\label{im1}
\end{equation} 

where, $t^{\prime}$ is an extra coordinate in expanded Hilbert space and $\Phi(x,t^{\prime},t)$ represents the solution of TDSE obtained from $(t,t^{\prime})$ method,

\begin{equation}
\iota\hbar \frac{\partial}{\partial t} \Phi(x,t^{\prime},t)=\mathcal{H}(x,t^{\prime})\Phi(x,t^{\prime},t).
\label{im2}
\end{equation} 

\indent $\mathcal{H}(x,t^{\prime})$ operator is of the form:

\begin{equation}
\mathcal{H}(x,t^{\prime})=H(x,t^{\prime})-\iota\hbar\frac{\partial}{\partial t^{\prime}} .
\label{im3}
\end{equation}

\indent As $\mathcal{H}(x,t^{\prime})$ is time independent, the time-dependent solution would be:

\begin{equation}
\Phi(x,t^{\prime},t)=\hat{U}(x,t^{\prime},t_{0}\rightarrow t)\Psi(x,t_{0}),
\label{im4}
\end{equation}

with,

\begin{equation}
\hat{U}(x,t^{\prime},t_{0} \rightarrow t)=e^{-\frac{\iota}{\hbar}\mathcal{H}(x,t^{\prime})(t-t_{0})}.
\label{im5}
\end{equation}

\indent The wave function $\Phi(x,t^{\prime},t)$ provides detailed dynamical information of the system under study. A finite grid (of sampling points in the $x$ and $t^{\prime}$ space) is used to define the wave function in such away that at the boundaries of the grid the wave function's amplitude is periodic or exponentially small . Periodic boundary conditions are normally taken for the $t^{\prime}$. The propagation method depends on an iterative scheme developed using the time independent Hamiltonian, $\mathcal{H}(x,t^{\prime})$ . The starting point is to carry out the operation of $\mathcal{H}(x,t^{\prime})$ on  $\psi(x,t^{\prime},t)$, i.e.

\begin{equation}
\phi(x,t^{\prime},t)=\mathcal{H}(x,t^{\prime})\psi(x,t^{\prime},t). 
\end{equation} 

\indent The Operator  $\mathcal{H}(x,t^{\prime})$ is regrouped in coordinate and momentum space as:

\begin{equation}
{\mathcal{H}(x,t^{\prime})}=V_{0}(x)+V_{t}(x,t^{\prime})-\frac{\hbar^2}{2m}\frac{\partial^2}{\partial x^2}-\iota\hbar\frac{\partial}{\partial t^{\prime}} ,
\end{equation} 

where $V_{t}$ represents the time dependent potential term  of the Hamiltonian.
The potential part and momentum part are then calculated at each grid points. However  momentum part is handled using fast Fourier transform algorithm [\ref{DKR},\ref{RK}].
\subsubsection{Split Operator Method}

In this method, time evolution operator  $U=e^{-\iota Ht/m}$ is approximated as product of kinetic and potential terms. Consider that operator $'U'$ is expressed as product of $'M'$ propagators in interval $[0,t]$ with time steps $\Delta t/M$, 

\begin{equation}
U(t,0)=\displaystyle\prod_{m=1}^{M} e^{(-H_m \Delta t/ \hbar)} ,
\label{sp1}
\end{equation}

where, for sufficiently small time step, $H_m=H(m \Delta t)$ can be considered as constant without any loss of generality.

By applying Zassenhaus formula [\ref{Tannor}]:

\begin{equation}
e^{A+B}= e^{A}e^{B} e^{-[A,B]/2} .... ,
\end{equation}

each propagator in equation (\ref{sp1}) is approximated as :

\begin{equation}
e^{-\iota H \Delta t/ \hbar}= e^{-\iota T \Delta t/\hbar}e^{-\iota V \Delta t/\hbar}+ \mathcal{O}(\Delta t^{2}) ,
\end{equation}

In above equation error associated with $ \Delta t^{2}$ depends on commutator [T,V]. However in case of symmetric splitting the errors associated with $\Delta t^{2}$ are eliminated and propagator is estimated as:

\begin{eqnarray}
 e^{-\iota H \Delta t / \hbar}  \approx  \left( e^{-\iota T \Delta t/\hbar}e^{-\iota V \Delta t/\hbar} \right) \left(  e^{-\iota V \Delta t/\hbar}e^{-\iota T \Delta t/\hbar} \right)+ \mathcal{O}(\Delta t^{3}) \nonumber \\
 =  e^{-\iota V \Delta t/2 \hbar}  e^{-\iota T \Delta t/2 \hbar}+\mathcal{O} (\Delta t^{3}) .
\end{eqnarray}

Hence, in order to evaluate time evolution propagator we need to evaluate two exponential matrices. Exponential potential matrix is non diagonal in nature and is evaluated evaluated using fast Fourier transform (FFT).
Hence, in the split-operator FFT method, at  each time step ∆t, we need to follow this order  : (I) Write the wavefunction, then apply the first potential factor of time evolution operator,further perform an FFT (in the momentum representation) after this apply
the kinetic factor then perform an inverse FFT (back to the position space). Finally apply the second potential factor of time evolution operator.
The propagation of the wavefunction in each time step depends on the its value in the previous time step at all points of the grid. More detail of this method and its applications are discudded in  references [\ref{Feit},\ref{Bandrauk}].

\subsection{Importance of Pulse Shapes in Rotational Dynamics}

 Ultrshort shaped laser pulses of durations ranging from nanoseconds to attoseconds  [\ref{FK1r}-\ref{IA1r}]  have been used to stabilise the target states of atomic and molecular  systems. Ultrshort pulses with desirable pulse shapes, pulse duration, strength, spectral widths  and repetition rates can be designed and generated by various techniques [\ref{JC1r}-\ref{TS1r}].
Appropriately shaped waveforms are also crucial for  many applications  in remote sensing, signal processing, atomic and molecular spectroscopy in form of  coded signals[\ref{Weiner}]. 
Different pulse shapes exploited in literature are discussed in detail in this section and their temporal profile is presented in figure 2.

\subsubsection{Half Cycle }

The commonly used HCP profiles in the literature  are of the form:

{\bf{Gaussian(G) temporal profile HCP}}:
\begin{equation}
f(t)=\exp[-\frac{t^{2}}{{\tau}^{2}}],
\end{equation}

where, time parameter,  $\tau$, signifies the pulse width.

{\bf{Sine-square($sin^{2}$) temporal profile HCP}}:

\begin{equation}
f(t) = \left\{
                 \begin{array}{l l l}
                      sin^{2}(\frac{\pi t}{\tau}) &,& 0<t<\tau \\ 
                         0 &,& otherwise
                 \end{array}
         \right.
\label{sinsqp}
\end{equation}	

{\bf{Strongly asymmetric HCP}}

 The asymmetric mono-cycle pulses have sharp tail of positive polarity and  a smooth tail with negative polarity. Duration of sharp tail is larger than smooth tail. Hence dynamics of the system exposed to these kind of asymmetric pulse is completely driven by the sharp tail. The mono cycle asymmetric pulse is called as full pulse, but sharp tail is called as half cycle pulse (HCP), while smooth tail, just called as tail, because it has negligible contribution towards the dynamics of the system.
Experimentally You {\it{et al.}} [\ref{DY1r}] generated  asymmetric HCP pulse in 1992. Recently, a theoretical model to generate an  attosecond  half  cycle pulses is reported in reference [\ref{JXu}].
 
Dynamics of the system exposed to HCPs is qualitatively very much different from  dynamics of the system in Symmetric (or nearly symmetric ) laser pulses . This is attributed to asymmetric nature of HCPs. Crucial point about HCPs is that a HCP delivers a non zero momentum to the system which is slightly decreased by opposite contribution from tail. This total momentum transfer to the system decreases slowly and prolongs for a duration much smaller than  duration of the asymmetric full pulse. If the characteristic period of rotation of the molecule is much longer than HCP duration molecular system can be studied under impulsive region. Instantaneous
momentum transferred to the system in impulsive regime  opens new dimensions to be explored in rotational dynamics. Recently rotational dynamics of polar molecules in impulsive region using subcycle unipolar pulses have been reported by [\ref{Arkhipov}].
 
Seideman has summarised studies done by his group on rotational,orientation and alignment dynamics of various molecular systems using half cycle pulses [\ref{TS1}] 
Recently, THz HCPs and  HCPs trains have been employed in non-adiabatic regime for orientation of polar molecule [\ref{CM1r},\ref{NEH},\ref{AM2r}].  
Dion etal [\ref{Dion}] have used $sin^2$ HCP mentioned above is used to study the orientation dynamics of $LiCl$ molecule using rigid rotor model. They have shown  that orientation in such kind of laser pulses is more sensitive to time integrated amplitude of interacting field than shape or its rising time. Molecular orientation using  half cycle pulses [\ref{AM2r}] and  combination of HCP pulse with a series of femtosecond pulses are studied theoretically [\ref{NEM}]. To orient molecules effectively half cycle pulses have been combined with delayed  laser pulse[\ref{EIR},\ref{EIS}}]. However by combining  HCP with nonresonant laser enhanced  strong field free orientation can be obtained [\ref{DDaems}].Other studies done in impulsive region to attain field free orientations using THs HCPs are [\ref{Dion1}, \ref{Machholm}].

\subsubsection{Single Cycle Pulse}

In addition to HCPs ,single cycle pulse is also important to explore rotational dynamics of the molecule. In this  light pulse  electric field have single oscillation cycle. The Guassian temporal profile of the field for single cycle pulse has the form: 

\begin{equation}
f(t)=\frac{t}{\tau} \exp[-\frac{t^{2}}{\tau^{2}}]
\end{equation}

When Single  cycle THz pulses  of zero area interacts resonantly  with rotating OCS molecules,  significant field free orientation  and alignment has been noticed experimentally[\ref{SFleischer}]. Jet cooled HBr molecules has been oriented using THz cycle pulses, as electric field is increased further degree of orientation can be further increased [\ref{KKT}]. Using  symmetric single cycle THz pulses, significant nonzero orientation over a rotational period has been obtained because of interference of time evolved eigenstates [\ref{JO3}]. Optical imaging technique for molecule rotor has been developed recently using alignment and antialignment obtained due to excitation caused by single linearly polarised pulses [\ref{JB}]

\subsubsection{Few Cycle pulses}

 Few cycle pulses are represented by following  theoretical models : 

{\bf{(a) Harmonic with a Gaussian envelope:}} 

The temporal profile of this pulse is:

\begin{equation}
f(t)= \exp[-\frac{t^{2}}{\tau^{2}}]cos(\omega t+ \phi),
\end{equation}

where, $\omega$ refers to the central frequency and $\tau$ is temporal width of envelope and $\phi$ is carrier-envelope phase.

{\bf{(b)Polynomial with Gaussian envelope:}} 

The model for few-cycle pulses is given by:

\begin{equation}
f(t)= \exp[-\frac{t^{2}}{\tau^{2}}]P(\frac{t}{\tau}),
\end{equation}
where P(x) implies a polynomial [\ref{Para}].

The pulses produced from the Er:fibre technology[\ref{Brid}], forms a realistic few-cycle pulses in the frequency domain [\ref{Mosk}]. Field free molecular orientation of LiH and LiCl molecules have been studied using Thz few cycle pulses (consisting of train of HCPs). This strategy is capable of generating  high degree of orientation and also overcomes the limitations  offered by HCPs[\ref{CCun}].

\subsubsection{Square Pulses(SQP)}

\indent In contrast to Gaussian pulses, the square pulses provide a sudden \lq\lq kick \rq\rq to the molecule and cause the rotational excitation to higher rotational states. The form of the square profile is as follows:
\begin{equation}
f(t) = \left\{
                 \begin{array}{l l l}
                      1 &,& 0<t<\tau \\ 
                         0 &,& otherwise.
                 \end{array}
         \right.
\label{sqpls}
\end{equation}	

In comparison to Gaussian pulse, non-resonant square pulses can be used in both adiabatic and non adiabatic rotational excitation to obtain field free molecular orientation [\ref{Shu-Wu}].
\indent The square pulses are obtained by tailoring the ultra-short laser pulse in the frequency domain [\ref{Weiner},\ref{AMS}]. 
Zhang {\it{et al.}} [\ref{Zhang}] have observed that square pulse is more rich than conventional Gaussian pulse in aligning the molecules. They have obtained field free molecular alignment in both adiabatic and nonadibatic regimes with nonresonant square laser pulse. (See this reference again try to write it in more elegant way)

\subsubsection{Ramped Pulses}

\indent Recently, it has been made possible to obtain the ramped pulses by tailoring the ultra-short laser pulses in the frequency domain [\ref{AMS}]. These laser pulses can be experimentally realized by the ultra-short pulse shaping method. The present technique is quite feasible and can be applied to various polar molecules. The form of a ramped pulse profile is as follows:

\begin{equation}
  f(t)= \left\{  \begin{array}{ll}
          0  & \mbox         {if  $ t \leq t_{0} $} ; \\
          sin^2[\frac{\pi}{2} (\frac{t-t_0}{t_1-t_0})] & \mbox  {if  $t_0 < t < t_1$} ; \\
          1  & \mbox         { if $t_1 < t < t_2$} ;\\
          sin^2[\frac{\pi}{2} (\frac{t_3-t}{t_3-t_2})] & \mbox  {if $ t_2 \leq t \leq t_3$};\\
	   0 & \mbox  {if $t \geq t_3$}. \end{array} \right. 
\label{ramppls}
\end{equation}

\subsubsection{Pulse Train}

The degree of molecular alignment obtained by using single laser pulse strongly  depends on the laser pulse  intensity. There is a maximum intensity beyond which application of laser will ionize  the molecule and it will also destroy the alignment obtained. In order to overcome the limit the  imposed by the  maximum intensity, multiple pulse method is used to achieve improved degree of alignment and avoid destruction effects in molecules [\ref{ML}-\ref{CZB}]. In this multiple pulse strategy, one or more additional pulses are applied after the certain delay of first aligning pulse. 

 \indent Initial studies to attain high degree of alignment using multiple pulse strategy  are [\ref{ISA},\ref{MLI}].    A pulse train  method is  used in [\ref{ISA}] to achieve enhanced alignment however in [\ref{MLI}], an optimal control theory,  based on  two and three pulses is discussed. \\
 
 In two pulse alignment strategy ,the alignment created by first aligning pulse can be suppressed or enhanced by second laser pulse [\ref{KFLEE}-\ref{GCH}].  The degree of suppression or enhancement depends on laser pulse shape and also on the delay time at which it is applied after the second pulse. Population distribution in various  molecular rotational states, annihilation or enhancement of alignment are strongly influenced by time delay between initial pulse and delayed pulse [\ref{CGY}-\ref{CGH1}]. Meijer etal [\ref{Meijer}] have reported influence of time delay on oscillatory behaviour of rotational states.
  
In non-adiabatic alignment, degree of alignment can be enhanced by applying more than one consecutive and separate laser pulses. Hence we get enhanced alignment in field free conditions. However  this is not the case with adiabatic alignment and alignment gets wash away as laser field is switched off.  Studies done in references [\ref{MLI},\ref{MLI}-\ref{SG2}] deals with three pulse mechanism where pulse shapes, pulse intensities  and pulse separation are varied systematically to get  field free alignment. Experimentally impulsive alignment for $N_{2}$ molecule  with eight consecutive identical laser pulses has been reported [\ref{JPC2}]. In  a pulse train of eight pulses separated by rotational period enhanced alignment has been noticed [\ref{JPC2},\ref{CB2}].

Pulse train methods to get enhanced alignment have been initially applied on linear and symmetric top molecules. In these cases well defined periodic revivals of alignment  have been noticed. However  in asymmetric molecules the rotational energy levels have irregular spacing , due to this rotational wavepacket reshaping is distorted and the alignment revivals are not periodic in nature[\ref{AR2}]. Alignment of asymmetric top molecules in one dimension and two dimension are reported in experimental studies done in references[\ref{AR2}-\ref{LH2}]. These experimental studies are based on two pulse alignment strategy.However three dimensional alignment for asymmetric top molecule using multi pulse strategy have been reported in [\ref{KFL2} and[\ref{ISA2}].  Multi pulse method used in reference [\ref{ISA2}] proposes to make use of pair of four pulses. First train of four pulses align the molecule and the second train of well separated pulses is applied in such a manner that molecule gets additional kick at the time when alignment due to previous pulse is maximum and hence there is continuous increase in the alignment. 

\indent It is quit challenging to orient the molecule than to align them  in field free conditions.In order to avoid the destructive effects caused by  a weak DC field many  schemes like  two-color shaped laser pulse [\ref{MMura}] ,two-color phase-locked laser pulse [\ref{Oda}-\ref{MSpa}], and multi-color laser pulse [\ref{SZha1}]have been used to achieve the orientation.  In literature the molecular orientation has also been achieved by a half-cycle THz laser pulses and few cycle pulses [\ref{CMDi},\ref{ATya}]. However, it is quite difficult to generate half cycle pulse experimentally[\ref{JRau}]. The maximum  degree of molecular orientation achieved by a single-pulse depends on temporal duration and pulse peak intensity . Enhanced field-free molecular orientation can be obtained from  a two-color shaped laser pulse method and a time-delayed THz laser pulse train method [\ref{HaiP}]. 

\begin{figure}
\vskip 10.5cm
\hskip 1.2cm
\includegraphics{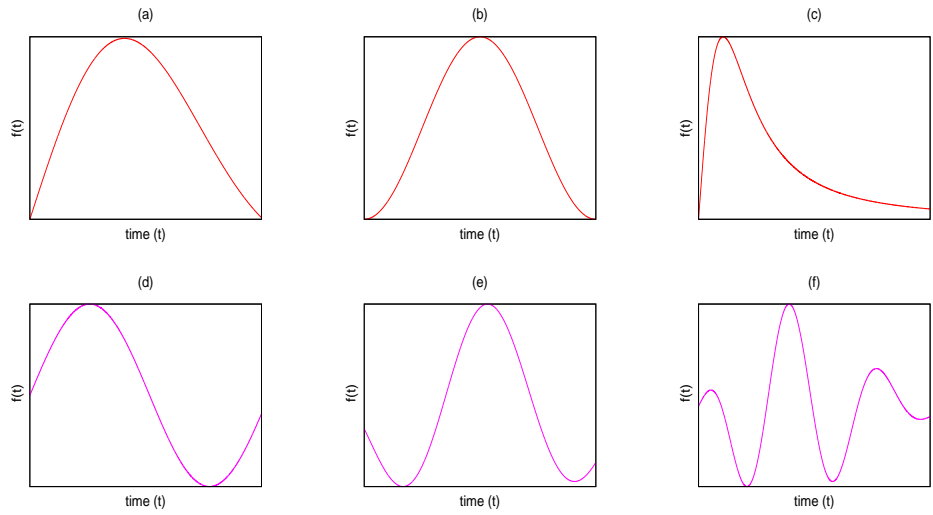}
\caption{The different temporal shapes $f(t)$ of broadband ultra-short light pulses   (a).  Gaussian profile, (b). Sine-square profile, (c). Strongly asymmetric HCP as (d). Single cycle pulse   (e). Harmonic few cycle pulse with Gaussian envelope, (f). Polynomial few cycle pulse with Gaussian envelope }
\label{pulse}
\end{figure}

\section{Various Approaches to Control Molecular Rotation}

\indent Controlling the interactions between light and matter has been a long standing aim in atomic and molecular physics. As already discussed, there exists a number of ways through which we can control the degree of orientation and alignment. Here, in the present section we would discuss few of the techniques to control molecular orientation and alignment in detail.  

\subsection{Orientation by Static Electric Field in Combination with Delayed Pulses}

\indent The interaction of a polar molecule with the simultaneous presence of three electric fields viz. continuous static field, second HCP with different pulse duraton and a strong field generated by the delayed zero area pulse/ultra-short HCP is [\ref{Arya}]:\\

\begin{equation}
H(t)=B\hat{J^{2}}+V_{s}+V_{E}(\theta,t)+V_{Z}(\theta^{\prime},t).
\end{equation}
  
where, $B$ refers to the rotational constant of the molecule and $\hat{J^{2}}$ signifies the squared angular momentum operator. The terms $V_{s}$, $V_{E}$ and $V_{Z}$ are the interaction potentials. The first interaction term is  $V_{s}=\mu_{0}E_{1}$, where, $\mu_{0}$ is the permanent  dipole moment of the molecule along the internuclear axis and $E_{1}$ is the static field amplitude. The second term represents the interaction potential with  HCP  laser field and has the form:

 \begin{equation}
V_{E}(\theta,t)=\mu_{0}E_{2}(t)cos(\theta),
\end{equation} 
 
where, $\theta$ refers to the polar angle between the molecular axis and laser field. This angle defines the orientation of the molecule $w.r.t.$ the laser field. Also,

\begin{equation}
E_{2}=E_{0}f(t)sin(\omega t),
\end{equation}

with, $E_{0}$ as the electric field amplitude and f(t) is the pulse envelope given by equation (\ref{sinsqp}). The interaction, $V_{Z}(\theta^{\prime},t)$ is the interaction due to zero area pulse/ ultrashort HCP:

\begin{equation}
V_{Z}(\theta^{\prime},t)=\mu_{0}E_{3}(t)cos(\theta^{\prime},t)
\end{equation}

In case of zero area pulse, $E_{3}(t)=sin(\frac{\pi t}{T_{p}})$ and for ultrashort HCP,  $E_{3}(t)=sin^{2}(\frac{\pi t}{T_{p}})$ with $T_P$ as pulse duration . Also $\theta^{\prime}$ represents zenith angle between molecular axis and zero area pulse/ultra-short HCP.

Time evolution of transition probabilities for different states depend on static field strength and also on form of delayed pulse applied figure(3) in reference [\ref{Arya}]. Initial pulse is applied with the pulse width of $1.0 ps$, however delayed pulse is applied at delay time $t_{c}/2$, where $t_{c}$ denotes the rotational period of the molecule. For $LiCl$, $t_{c}=23.6 ps$. It is seen that as static field strength $E_s$ is increased the population distribution become more oscillatory in nature and  population of  higher rotational states increases. However some lower rotational states remain unaffected by static field variation, however there oscillatory behaviour is still dependent on static field strength. Zero area pulse induces dominating oscillatory behaviour  impact than ultra-short HCP on  transition probabilities.    

The orientation dynamics of dipolar molecules in delayed pulses shows interesting behaviour. It is noticed in  studies done in reference [\ref{UB}], (figure 7(b)), that  when molecule is exposed to only HCP, Orientation  increases with increase in pulse duration, however when HCP is applied in combination with delayed ultrshort HCP the dynamics gets reversed and maximum orientation is attained when initial HCP pulse is exposed for small duration. 

\subsection{Orientation by Ramped Pulses}

Nondiabatic rotational excitation dynamics and orientation play important role for many applications in the fields of chemistry and Physics [\ref{Charron}-\ref{Ghafur}] . Single pulse approach to stusdy nonadiabatic alignment  is practiced by many []. However when we apply laser pulses in different combinations NAREX changes appreciably and gives reasonable field free orientation [\ref{Renard}-\ref{Jortigoso},\ref{SVH},\ref{ISA},\ref{SZha},\ref{SZha1}]. 
 
 Urvashi {\it{et al.}} [\ref{UBV}]  have studied  NAREX using ramped pulses and combining ramped pulses with Gaussian pulses and  square pulses. They have also discussed impact of ramped pulses on the orientation of a polar $HBr$ molecule [\ref{UBV}].  Using rigid rotor approximation, the effective Hamiltonian for molecule with interacting laser field  is given by:
 
\begin{equation}
H(t)=B\hat{J^{2}}+V_{E}(\theta,t),
\end{equation}
	
Here, $B$ and $\hat{J^2}$ are the rotational constant and squared angular momentum operator, respectively. The interaction potential $V_{E}(\theta,t)$ is due to the laser field interaction  with permanent  dipole moment and is defined as:

\begin{equation}
V_{E}(\theta,t)=\mu_{0}E_{i}(t)cos(\theta).
\end{equation}
 
Here, $\theta$ is the polar angle between the laser field  direction and the molecular axis which defines the orientation of molecule with respect to laser field. The laser field, $E_{i}$ for the present case is :

\begin{equation}
E_{i}(t)=\sum_{i=1}^{2} E_{0i}(t)sin(\omega_{i}t+\phi_{i}).
\end{equation}

The quantities $\phi$ and $\omega$ are the phase difference and frequency of the pulse. Also;

\begin{equation}
E_{0i}=E_{01}f(t).
\end{equation}

The quantity, $E_{01}$ is the electric field amplitude and $f(t)$ is the envelope for different pulses. In order to make energy spacing between rotational states comparable with frequency of the delayed pulse, it is necessary to first apply the initial laser pulse for exciting the molecule before applying delayed pulse. Envelope function $f(t)$ for ramp pulse, square pulse and Gaussian pulse is defined by equations (\ref{ramppls}) ,(\ref{sqpls}) and (\ref{sinsqp}), respectively.   

Maximum value of the orientation vary appreciably  with delay time of the delayed pulse [\ref{UBV},figure(3)], for  pulse combinations: (i)ramp-ramp Pulses (ii) Gaussian-Gaussian pulses (iii) Gaussian-Ramp pulses combination, at temperatures $0K$ and $20 K$. Laser parameters taken are $E_{01}=500 KV/cm$ , Pulse duration taken is $0.3 ps$ and laser intensity  is $ 3.3 \times 10^{12}  W/cm^{2} $. Delayed pulse is applied after $2.1 ps$  For all combinations,  molecular orientation first increases linearly with pulse delay time  and then after attaining maximum it decreases. Maximum value of $<cos \theta>$ has been obtained for  Ramp-ramp pulse combination in comparison to other combinations. Maximum orientation also decreases with increase in temperature.

Population dynamics of various  states can be controlled by the laser intensity of the delayed pulse  and delay time at which the second pulse is sent (reference [\ref{UBV}],figure(5)). Second pulse is applied at the the delay of $2.1 ps$. Other parameters like field amplitude and pulse duration are $500kV/cm$ and $0.3ps$, respectively. For low values of laser field intensity of the delayed pulse,  lower rotational states are populated but as the intensity  increases, coupling between different rotational states takes place, which in turn enhances the  population in higher rotational states .

\subsection{Non-adiabatic Alignment by Combination of Orienting and Aligning Pulse}

In the model considered  in reference[\ref{UAV}], a polar molecule of moderate dipole moment, $\mu_{0}$ is allowed to interact with the orienting field and delayed aligning field (which is Sine Square(SS) or Square Pulse(SQP)).

\indent  Interaction Hamiltonian of this model is:
\begin{equation}
H(t)=B\hat{J^{2}}+V_{E}(\theta,t)+V_{Z}(\theta,t),
\end{equation}

with,  $B$ as rotational constant for HBr molecule and $\hat{J^{2}}$ represnts squared angular momentum operator. Also, $V_{E}(\theta,t)$ and $V_{Z}(\theta,t)$ represent the interaction potentials of the orienting pulse  and IRL  delayed aligning pulse, respectively. 
Interaction term $V_{E}(\theta,t$ has the form: 

\begin{equation}
V_{E}(\theta,t)=-\mu_{0}E(t)cos(\theta),
\end{equation}

 with, $E(t)=E_{0}f(t)$, where $E_{0}$ defines amplitude of the orienting pulse and $f(t)$ is the envelope function (eqn (\ref{env4})).
 
However aligning pulse interaction potential, $V_{Z}(\theta,t)$, is given by:

\begin{equation}
V_{Z}(\theta,t)=-\frac{1}{4}[E_{01}(t)]^{2}(\Delta \alpha cos^{2}\theta+\alpha_{\perp}),
\end{equation}

with, $E_{1}(t)= E_{01}f(t)$, here $E_{01}$ represents the amplitude of the delayed IRL pulse
and envelope function is given by the following expression:

\begin {equation}
f(t)= \frac{1+cosh(t_p/2t_f)}{[cosh((t-t_0)/t_f)+cosh(t_p/2 t_f)]}.
   \label{env4}
\end {equation}

where, $t_p$ is the pulse duration and  and $t_f$ is the front pulse duration. Shape of  pulse is dependent on $t_{f}$. When $t_f =0ps$, f(t) expresses  SQR pulse and for $t_f \geq t_p$, f(t) represents SS pulse. 

As $t_f$ changes from $0.01 ps$ to $0.1 ps$ the shape of the pulse evolves from SQP to SS (Gaussian Pulse) as shown in figure(1)of reference [\ref{UAV}]. When shape changing delayed pulse is applied after the initial pulse of a fixed shape by keeping  maximum field strength and interaction time fixed, interesting after pulse rotational dynamics is noticed. The alignment depends on  the extremities  of $\langle cos^{2}\theta \rangle$.

Shape of orienting pulse plays an important role in rotational population dynamics as shown in figures (2) and (3) in reference [\ref{UAV}]). When initial orienting pulse  of Gaussian (SS) shape with pulse duration of $0.1 ps$  and delayed pulse (of SQR or SS shape with delay time $1.7 ps$ and laser intensity $5 \times 10^{3}$) is applied  transfer of population to even states occur. Population transfer to different rotational states  also depend on shape of delayed pulse applied. Comparison of population transfer obtained with different shapes of delayed pulse shows that    SS shape (Gaussian pulse) promotes population transfer to lower rotational states ($J\le 4$) as compare to SS pulse , however  SQP shape of the delayed pulse promotes population transfer to higher rotational states (for $J > 4$). When orienting initial pulse used is of SQP shape instead of SS shap,e population transfer to odd rotational states occur. Hence SQP orienting pulse increases the population in odd rotational state , this in turn enhances the degree of alignment to an appreciably amount. Hence choice of  SQP as initial pulse  offers  a benefit of interesting rotational dynamics.   

Dependence of Nonadibatic alignment dynamics  on the delay time ( at which delayed pulse is applied) and temperature, is also discussed in reference [\ref{UAV}] in figure(9). SS shaped initial pulse  with SQP delayed pulse having pulse durations of $0.4 ps$ and $0.2 ps$, respectively ,   gives ${\langle cos^{2}\theta \rangle}_{max}$  optimum at  delay time $T_{d}=1.1 ps$. Intensity of delayed pulse is $5 \times 10^{13} W/cm_{2}$. Adding SQP as delayed pulse to SS initial  pulse imparts a kick to molecule farther  in the same direction as initial pulse  which in turn further increases degree of alignment. Temperature also plays important role in the alignment dynamics of the molecule. At $T=0 K$, there is no population in higher rotational states and all population resides in ground state, as a result no alignment takes place at $T=0 K$. However with increasing temperature population transfer to higher rotational states take place  and at $T=1 K$ it is seen to be maximum. Further increase in temperature causes decrease in population which starts destroying alignment (or decreases the alignment).

\subsection{Non-adiabatic 2-D Alignment by Shaped Laser Pulses}

Alignment of molecules under field free conditions depends on many parameters like intensity of laser pulses, field strengths, pulse width, pulse duration and shape of the pulses. In addition to this state of polarization of electric fields has also important role in one dimensional (1D), two dimensional (2D) or three dimensional (3D) alignment ([\ref{ADV}, \ref{Rouz}, \ref{Daem}, \ref{Chh}, \ref{ML2},\ref{Leo}]).   

\indent The elliptically polarized laser field used to attain '2D' alignment has the form      [\ref{ADV},\ref{Rouz},\ref{Daem}]:

\begin{equation}
\vec{E(t)}=E_{0}(t)[\vec{e_{x}}a cos(\omega t)+ \vec{e_{y}} b sin(\omega t)],
\label{epl}
\end{equation}

 where, $E_{0}(t)$ and $\omega$, are the amplitude and optical frequency of the laser field, respectively. Also,\, $'a'$ is the half-axis of the ellipse (along the laboratory, x-axis) whereas $'b'$ corresponds to half axis of the ellipse along the y-axis, with $a^{2}+b^{2}=1$ and $a > b$. In above equation, $E_{0}(t)=E_{0}f(t)$, with $E_{0}$, as the  electric field amplitude and $f(t)$ represents the pulse envelope. Envelope function for half cycle pulse(HCP) and Square pulse, are defined by equations (\ref{sinsqp}) and  (\ref{sqpls}), respectively.


 The '2D' alignment parameters are obtained as follows:
 
\begin{equation}
\langle cos^{2} \theta_{z} \rangle(t) \equiv \langle \psi(\theta_{z},\phi_{z};t)\mid cos^{2}\theta_{z} \mid \psi(\theta_{z},\phi_{z};t)\rangle ,
\end{equation}

\begin{equation}
\langle cos^{2} \theta_{x} \rangle(t) \equiv \langle cos^{2}\phi_{z} sin^{2}\theta_{z}\rangle (t),
\end{equation}

and

\begin{equation}
\langle cos^{2} \theta_{y} \rangle(t) \equiv \langle sin^{2}\phi_{z} sin^{2}\theta_{z}\rangle (t),
\end{equation}

and

where, $\theta_{x}(\theta_{y})$ represents  polar angle {\it{w.r.t.}} the x-axis (y-axis).  By virtue of the relation [\ref{Daem}]:

\begin{equation}
\Sigma_{i=x,y,z}{\langle cos^{2}\theta_{i}} \rangle (t)=1,
\end{equation}

 the alignment variation can be measured by any pair of observables out of $[\langle cos^{2} \theta_{i} \rangle (t),i=x,y,z]$. Perfect alignment is attained for $\langle cos^{2} \theta \rangle(t)$ $\epsilon$[0,1] acquires its extreme values.
 
When selected molecule is subjected to HCP-SQP and HCP-HCP pairs of initial and delayed elliptically polarized pulses,  interesting  NAREX dynamics occurs as shown in figure(2) of reference [\ref{ADV}]. Elliptical parameters are $a_{i}=a_{d}=0.8$, here $a_{i}$ is for initial pulse and $a_{d}$ is for delayed pulse. Delayed pulse is sent at  $1.8 ps$, also pulse duration for both the pulses is $0.3 ps$. Initial pulses and delayed pulses have same intensity  $5 \times 10^{14} W/cm^{2}$. In HCP-SQP combination SQP pulse have sharp rising and falling edges which causes increase in population of higher rotational states. However this kind of trend is not observed in case of HCP-HCP pair of pulses. This observation reflects that shape of delayed pulse is an important  parameter to annihilate or enhance the population of the higher rotational states. Increase of population in higher rotational states further suppresses the degree of alignment. Hence the suitable choice of delayed laser pulse  helps to obtain desired alignment along x-axis and z-axis. Delayed SQP pulse suppresses alignment to larger extent along z-direction than delayed HCP. In general HCP-HCP combination gives higher degree of alignment than SQP-SQP pair of initial and delayed pulses.
    
 2D alignment analysis for four combinations of initial and delayed pulses {\it{viz.}} SQP-HCP, SQP-SQP, HCP-HCP and HCP-SQP combinations shows that dominance of alignment in a particular direction depends on shape of the delayed pulse (figure 4 of [\ref{ADV}]). When delayed elliptic parameter $a_d$ is varied from $0 $ to $1$, SQP-HCP and HCP-HCP pairs provide higher alignment along x-direction whereas SQP-SQP and HCP-SQP pairs support  higher alignment along z-direction. Hence whatever is the nature of initial pulse, maximum alignment direction depends on the the shape of delayed pulse. When HCP delayed pulse is applied, interaction time of molecule and pulse increases because of gradual rising or falling edges, hence molecule is aligned more towards the x-direction. However when SQP is applied as delayed pulse, its sharp rising or falling edges imparts a strong kick, which causes molecule to align more towards z-direction. For HCP-HCP combination maximum  2DA in z-direction dominates at low values of delay time ( figure(6),[\ref{ADV}]).

\subsection{Orientation and Alignment by Combination of Orienting Pulse and Aligning Pulse Train}

At higher intensities, some nonlinear effects [\ref{Chot}] hinders the alignment of the molecules. When ionization saturation occurs, molecule gets ionized. However, by using pulse train  method to align or orient a molecule this situation can be avoided because this method distributes the energy of a single pulse over many sub pulses by conserving the sub pulse duration . This is also one of the efficient method to enhance the degree of alignment in field free conditions [\ref{ATya},\ref{AA22}].

The total Hamiltonian for molecule field interaction in rigid rotor approximation is:
\begin {equation}
  H(t)= B\hat{J^2} + V_{E}(\theta,t) + V_{Z}(\theta,t), 
  \end {equation}
  
where, B  and $\hat{J^{2}}$, symbols have usual meanings as described earlier. Terms,  $V_E(\theta, t)$ and $V_Z(\theta, t)$, represent the interaction potential of initial the orienting field and  delayed IRL pulse train,  respectively.  Both fields are in field polarization direction. 

The interaction, $V_{E}$ is defined as:
 
\begin{equation}
   V_{E}(\theta,t)=-\mu_{0}E(t)cos(\theta),
\end{equation}

with, $E(t)=E_{0} f(t)$, $E_{0}$ is amplitude of the orienting field pulse and f(t) is the envelope of the pulse. Also, $\theta$ is the angle between the orienting pulse the molecular axis and $\mu_{0}$ stands for the permanent  dipole moment of the molecule along the internuclear axis.

The term, $V_{Z}(\theta, t)$ is an interaction due to the delayed aligning pulse train and has the form:

\begin{equation}
   V_{z}(\theta,t)=-1/4[E_{1}(t)]^{2}({\Delta\alpha cos^2(\theta)}+\alpha_{\bot}),
\end{equation}

where, $\Delta\alpha = \alpha_{\|}-\alpha_{\bot}$  is the difference between the parallel ($\|$) and perpendicular ($\bot$) components of polarizability tensor of molecule. Also, $E_{1}(t)= E_{01}f(t)$, with $E_{01}$, as amplitude of delayed pulse and f(t) is the envelope of the pulse.

Pulse train width have noticeable impact on the population of the higher rotational states and alignment dynamics changes accordingly, as noticed in figure (1) of reference [\ref{AA22}]. Initial pulse which orients the molecule is $Sin^{2} (HCP)$ and has pulse width, $0.2 ps$. Delayed aligning IRL pulse train of five sub-pulses  ( $Sin^{2} (HCP)$ )is applied at delay time, $1.9 ps$. For $Sin^{2}-Sin^{2}$ pulse combination of initial (orienting) and delayed (aligning) pulses  the smaller pulse train width(PTW) of $0.1 ps$ has more prominent impact on  population and alignment  dynamics, than  pulse train widths $0.2 ps$ and $0.3 ps$. Similar results are noticed for $SQP-SQP$ pulse combination of initial (orienting) and delayed (aligning) pulse  with pulse train widths $0.1 ps$, $0.2 ps$ and $0.3 ps$. Hence smaller pulse width (i.e. $0.1 ps$) is more suitable to enhance the population of the higher rotational states. In $Sin^{2}-Sin^{2}$ pulse combination  maximum population of 0.6 is achieved for $J=2$ state, however in  SQP-SQP combination of orienting and aligning pulse  maximum population of 0.4   occurs for $J=4$ rotational state. In conclusion,  population of higher rotational states and as well as alignment dynamics depends on pulse train width  and also on the shape of the delayed and orienting pulse. SQP delayed pulse is responsible for transferring the population to higher rotational states. However degree of alignment is more in  $Sin^{2}-Sin^{2}$ pulse combination(of initial and delayed pulse) than SQP-SQP combination. This is attributed to higher population of J=2 rotational state (in $sin^{2}-Sin^{2}$ pulse combination ) than J=4 rotational state (in SQP-SQP combination).

Maximum alignment at three different values of delay time {\it{i.e.}}, $T_{d}=1.8 ps, 1.9 ps, 2.0 ps$ ,for three combinations;  (i) $Sin^{2}-Sin^{2}$ (ii) $Sin^{2}-SQP$  and (iii)SQP-SQP pulse  combinations, of orienting and aligning pulse has been shown in figure(2) of reference [\ref{AA22}]. Orienting pulse width (OPW) is $0.2 ps$ and pulse train width (PTW) for delayed pulse is $0.1 ps$ for all combinations. When initial pulse is of  $Sin^{2}$ shape it causes maximum population transfer in lower J=2 state, further application of  $Sin^{2}$ delayed pulse train enhances the population of the rotational state $J=2$ and hence causes maximum alignment. However, when delayed SQP train is applied along with the  $Sin^{2}$ orienting puls,e it does not further enhances the population of the lower rotational state, but promotes the population of higher rotational states, which causes slight decrease in the maximum degree of alignment. Also, for $SQP-SQP$ pulse combination population transfer occurs for higher rotational states than lower rotational states and hence low alignment is obtained in this case for all three delay times.  

Pulse shapes of orienting and aligning  pulse trains are important criteria in selecting the number of pulses required to transfer the population in desired rotational states [\ref{ATya}]. When $Sin^{2}-Sin^{2}$ and $Square-Square$, pulse combinations of orienting and aligning pulse are used to study population dynamics they show interesting dependence on number of pulses (figure(3),reference [\ref{AA22}]). For $Sin^{2}-Sin^{2}$ combination the appreciable  population in higher  rotational states  occurs for large number of pulse trains, however for $Square-Square$ pulse combination appreciable degree of alignment occurs for intermediate number of pulse trains.

Population dynamics of a particular rotational state depends on laser field strength of the orienting pulse and intensity of the aligning pulses(figure(4),reference[\ref{AA22}]). High amplitude orienting pulses and high intensity aligning pulses increase the population in high rotational states. At lower laser field strength of initial pulse only the lower rotational state, J=2 is populated. But as field strength of orienting pulse or intensity of the delayed pulse is increased, coupling between various rotational states occurs and it enhances population of higher rotational states.

\subsection{Near-adiabatic Orientation and Alignment by Mixed Field}

Mixed field dynamics has been used in literature to understand the behaviour of molecules or their ensembles. Stereo-chemical properties of hydroxyl (OH), free radical are studied analytically, numerically in combined electric and magnetic fields [\ref{Maed}-\ref{Mari}]. These studies also explore the effect of combined fields on energy levels. A moderate and intense nanosecond laser pulse is efficient to align the molecules adiabatically. However  when such nonresonant laser pulses are combined with weak  electrostatic  fields interesting nonadiabatic dynamics results in strong orientation and alignment, as noticed for OCS molecule  in  reference [\ref{Niel}]. Omiste {\it{et al.}} [\ref{Omis}] have also investigated theoretically, mixed field  orientation dynamics  for  linear polar molecules. They have suggested to use  rotational temperature below $0.7 K$ and strong static filed along with laser pulse, in order to achieve high degree of orientation in experiments. 

Mixed field dynamics of molecules in mixed field is not completely adiabatic, if the fields are parallel, a non-adiabatic population transfer of occurs  due to formation of quasidegenrate doublets in pendular states on increasing laser intensity, Hence orientation obtained is small as compare to adiabatic case. Also if the fields are non- parallel, ground states are highly oriented, but higher rotational states have moderate orientation values and some of higher states may occur as dark states [\ref{Omis1}]. In case of asymmetric rotors,  permanent dipole moment does not point towards any of polarizability axis. When such species is treated with an elliptically  polarised laser pulse, in combination with a weak electrostatic field (which does not coincides with any major or minor polarisation axis of the laser) aligns the molecule in  3D  and the most polarizable axis orients along major  polarization axis (1D orientation) [\ref{Hans}]. Field dressed system dynamics of  KCL  molecule is discussed  classically and quantum mechanically in detail [\ref{Aran}] . Also   orientation of  adsorbed molecule in conical well is analysed  in combined static electric and laser  field [\ref{Brij}].. 
 
 Moderate laser  fields tilted at an angle, when interact with molecule an interesting field dressed dynamics is noticed in adiabatic, nonadibatic and near adiabatic regimes [[\ref{Anjali3}]]. The dipole couplings due to two tilted laser fields have the following form:

\begin{equation}
H_{1}(\theta,t)=-\vec{\mu} . \vec{E_{1}}(t)= -\mu E_{1}f(t)cos(\theta),
\label{me1}
\end{equation}

and

\begin{equation}
H_{2}(\theta,t)=-\vec{\mu} . \vec{E_{2}}(t)= -\mu E_{2}f(t)cos(\theta_{L}),
\label{me2}
\end{equation}

where, 

\begin{equation}
cos \theta_{L}=cos \beta cos \theta + sin \beta sin \theta cos \phi.
\label{me3}
\end{equation}

In expressions, ($\ref{me2}$) and ($\ref{me3}$), $E_{2}(t)=E_{2} f(t)$ and $E_{1}(t)=E_1f(t)$, with $E_1$ and $E_2$, as field amplitudes of the laser pulses considered and f(t)is the envelope function. For  Sine Square pulse  and square pulse f(t) is  given by equations ($\ref{sinsqp}$)  and ($\ref{sqpls}$), respectively.

Double-pulse mechanism is an efficient technique to control rotational excitation dynamics and hence to get control on alignment of the system. When Sine square pulse is applied as both initial pulse and delayed pulse at delay time, $1.6 ps$. Initial pulse has pulse width of $ 1.0 ps$. Pulse width of the delayed pulse plays significant role in controlling the population. When pulse width of the delayed pulse is  near-adiabatic regime (i.e. 1.0 ps), appreciable degrees of orientation and alignment are obtained as compare to adiabatic and non-adiabatic regimes, with   delayed pulse widths of $2.5 ps$ and $0.2 ps$, respectively. Both the pulses have moderate field strength  of $ 250 kV/cm$. Delayed pulse, applied at tilt angle, $\beta= \pi/4$, cause splitting of rotational levels further into magnetic levels.  Moderate fields cause population transfer to only first few rotational levels than to higher rotational levels. Initial pulse excites the molecule and delayed pulse may  increase or decrease the population in lower or higher rotational levels depending on the interaction time of the pulse with the molecule. Kick mechanism of delayed pulse is strongest in near-adiabatic regime hence gives best orientation and alignment (figure(3), reference[\ref{Anjali3}]). 

When $Sin^2-Sin^2$ pulse combination is applied as initial and double pulse, higher degree of orientation  near-adiabatic regime is noticed, however, by  changing tilt angle it can be further modified (figure(4),reference [\ref{Anjali3}]). On comparing mixed field dynamics in near adiabatic regime for tilt angles, $\beta =0$, $\beta=pi/4$ and $\beta=\pi/2$, it is noticed that for parallel laser pulses population transfer is highest for $S3$ state and hence, high value of orientation and alignment. However, as  tilt angle, $\beta$ is increased from $\pi/4 $ to $\pi/2$, a decrease in orientation and alignment occurs due to decrease in population in $S3$ state. Actually, tilt angle divides the force,  which is being applied along z axis, hence causes decrease in population,orientation and alignment.    
  
Variations of maximum orientation, $\langle cos \theta \rangle$ and alignment, $\langle cos^{2} \theta \rangle$,  with delay time of delayed pulse at three different tilt angles; $\beta=0, \pi/4, \pi/2$, for $Sin^2-Sin^2$ pulse combination of initial and delayed pulse shows interesting behaviour in near-adiabatic and non-adiabatic region (as seen in figure(6) of reference [\ref{Anjali3}]). Both the pulses are considered under moderate field strength {\it{i.e.}} $250 kV/cm$. Initial and delayed pulses are having same pulse width of $1.0 ps$  For near-adiabatic interaction,when tilt angles are $\beta=0,\pi/4$  the orientation and alignment maxima are minimum and maximum  at   delay time near to $T_{rot}/2$ and  $T_{rot}$, respectively. Similar behaviour is observed in case of nonadibatic case. For $\beta=\pi/2$, in near-adiabatic case  maximum value of orientation or alignment becomes almost constant, however this behaviour is not noticed in nonadiabatic case. 

Population dependence of various rotational states on tilt angle for $Sin^2-Sin^2$ combination of initial and delayed pulses shows interesting behaviour in figure(9) of [\ref{Anjali3}]. Rotational states are designated as; $L1(J=0,M=0)$, $L2(J=1,M=-1)$, $L3(J=1,M=0)$, $L4(J=1,M=1)$, $L5(J=2,M=-2)$, $L6(J=2,M=-1)$, $L7(J=2,M=0)$, $L8(J=2,M=1)$ and $L9(J=2,M=2)$. 
When field strength of the delayed pulse is greater than the initial pulse ( {\it{i.e.}} $E_1:E_2=1:1.5$), population of the various rotational state is high as compare to case when field strengths of both the fields are same ({\it{i.e.}} $E_1:E_2=1:1$). However, for  individual rotational states  population transfer, with $\beta$ follows same trend for $E_1:E_2=1:1.5$ and  
$E_1:E_2=1:1$. Population of $L1$ and $L3$ states shows opposite behaviour with tilt angle, $\beta$. Initial pulse excite the molecule, but as delayed pulse is applied increase in  $\beta $,   transfers population of $L3$ state transfer to  $L1$ state, but this trend reverses after $\beta=0.5 radians$. However, $L2$ and $L4$ states have same population. Population transfer behaviour  with $\beta$, for L7  is opposite to that of L2 states. As tilt angle, $\beta $  is increased, population from higher rotational state L7 transfers to   lower rotational state L2, however after $\beta=0.5$ trend reverses. Hence by varying $\beta$ the population of different rotational levels can be controlled.

\subsection{Orientation by Delayed Elliptically Polarized Laser Pulses}

\indent  When the elliptically polarized laser pulses interact with a molecule, it generates rotational  wave-packets of longer durability for a particular rotational state. In  double pulse mechanism with some time delay between two pulses, initial  pulse creates a wave-packet (with $\Delta J=\pm 1$ and $\Delta M=\pm 1$), with first excited states, then further application  of  a  delayed (elliptically polarized) pulse leads to the interference of wave-packets which in turn  enhances or suppresses orientation parameter. 

In rigid-rotor approximation, the interaction Hamiltonian for molecule field interaction is given by   ( in atomic units, $\hbar$=$m_{e}$=$e$=1):

\begin{equation}
H(t)=B\hat{J}^{2}+V_{1}+V_{2} ,
\label{eq4}
\end{equation}

where, $B$ is molecule's the rotational constant, $\hat{J}^{2}$ is operator for the squared angular momentum. Also, $V_{1}$ and $V_{2}$, [\ref{Fang}]  are interactions of the molecule  with the initial and delayed elliptically polarised pulses (given by eqn(\ref{epl})). Both interaction terms are equal and given by:

\begin{eqnarray}
V_{I=1,2}&=&-\vec\mu_{0}.\vec E(t)\\
       &=&-[E_{0}(t)a\mu_{0}\sin\theta \cos\phi \cos(\omega t)+ E_{0}(t)b \mu_{0} \sin\theta \sin\phi \sin(\omega t)]\\
       &=&-[E_{0}g(t)a \mu_{0} \sin\theta \cos\phi \cos(\omega t)+ E_{0}g(t)b \mu_{0}\sin\theta \sin\phi \sin(\omega t)],
\label{eq5}
\end{eqnarray}

with, $\vec\mu_{0}$ as molecule's the permanent dipole moment. Here   $\theta$ , $\phi$ are  polar and Azimuthal angles  orthogonal to  ($x, y$) plane of the ellipse.

\indent The quantum dynamics of nonadibatic excitation can be explored by solving  time-dependent Schr\"odinger equation (TDSE) : 

\begin{equation}
\iota\frac{\partial{\psi(\theta;\phi;t)}}{\partial t}=H(t){\psi(\theta;\phi;t)}.
\label{eq6}
\end{equation}

The solution of above equation can be expressed as linear combination of unperturbed roatational eigenstates of the molecule and is given as :

\begin{equation}
\psi(\theta;\phi;t)=\sum_{J=0}^{J_{max}}\sum_{M=-J}^{J} C_{J,M}(t)\mid J,M \rangle \exp(\frac{-iE_{J}t}{\hbar}),
\label{eq7}
\end{equation}

Where, $C_{J,M}(t)$ is the expansion coefficient,  $E_{J}$ stands for the eigenenergy of the rotational state, $\mid J,M \rangle $. The dipole matrix elements [\ref{Fang}] are: 

\begin{eqnarray}
\langle J,M \mid \sin\theta \cos\phi \mid J-1,M-1 \rangle& =&-\iota\langle J,M \mid \sin\theta \sin \phi \mid J-1,M-1 \rangle \\ &=& -\frac{1}{2}\sqrt \frac{(J+M-1)(J+M)}{(2J-1)(2J+1)},\\
\langle J,M \mid \sin\theta \cos\phi \mid J-1,M+1 \rangle& =& \iota \langle J,M \mid \sin\theta \sin\phi \mid J-1,M+1 \rangle \\ &=& \frac{1}{2}\sqrt \frac{(J-M-1)(J-M)}{(2J-1)(2J+1)},\\
\langle J,M \mid \cos\theta \mid J-1,M \rangle &=& \sqrt \frac{(J+M)(J-M)}{(2J-1)(2J+1)}. 
\label{eq8}
\end{eqnarray}

The orientation in the $x$-axis, is  measured as:

\begin{equation}
\langle \cos \theta_{x} \rangle(t) \equiv \langle \psi(\theta;\phi;t)\mid \sin\theta \cos\phi \mid \psi(\theta;\phi;t)\rangle .
\label{eq9}
\end{equation}

where, $\theta_{x}$ is the  polar angle w.r.t. the $x$-axis. \\ 

Elliptically polarised pulses combination offers an effective mechanism to control rotational dynamics in non-dramatic regime. Initial pulse and delayed pulse with a delay time of 1.9 ps are applied to analyse the non-adiabatic dynamics. Initial pulse and delayed pulse have $a_{1}$ and $a_2$, as elliptical parameters, which are same  i.e. $a_1=a_2=0.5$. Both the pulses are of $Sin^2$ shape. Electric field strengths for both the pulses are $7.3\,MV/cm$.  On comparing, rotational dynamics at three different  pulse widths i.e. $0.1 ,0.2$ and $0.3$, it is noticed that probabilities  of higher rotational states increase and as a result orientation parameter also increases. Increase in pulse width causes constructive interference of the rotational wave-packets and this in turn, increases the orientation . On the other hand, small pulse width causes no interference and hence not any increase is observed in orientation parameter. Hence, $Sin^2$ elliptically pulses of large widths are much suitable to attain high degree of orientation, as noticed in figure(1) of reference [\ref{AAA22}). 

However when SQP pulse pair is used with all parameters same as above mentioned. Increase in orientation occurs for pulse widths $0.1 ps$ and $0.2 ps$. However, as pulse width is increased to $0.3 ps$, rotational probability of higher states does not increase and hence a decrease in orientation occurs. Hence SQP pulses of large widths are not suitable for getting higher degree of orientation as shown in figure(2) of [\ref{AAA22}].         

Elliptical parameters $a_1$ and $a_2$ also play important role in controlling orientation dynamics. When only initial pulse of $Sin^{2}$ shape   is applied and elliptical parameter $a_{1}$ is varied , maximum orientation also varies. As $a_1$ is increased, maximum orientation increases and it attains maxima for a particular $a_1$ value, further increase in $a_1$ after that causes the decrease in maximum orientation. When delayed pulse is applied at delay time of 1.9 ps after initial pulse due to kick mechanism, orientation parameter increases (than the case of single pulse). When  $a_2$ is varied, maximum orientation varies in  similar manner as in case of single pulse for $a_1$ variation (figure(7) in [\ref{AAA22}]).
     
Hence pulse shape, pulse width and elliptical parameters as well as field amplitudes are important parameters to suppress or enhance the rotational population in higher rotational states and hence  the orientation parameter.

\section{Collisional Rotational Excitation Dynamics}

Collisional processes in absence of laser fields has been investigated extensively both theoretically and experimentally [\ref{Levine}-\ref{Yang2}].Laser induced collisional processes have been explored since the beginning of 1970 and much progress has been obtained in theoretical work in following years [\ref{Gud}]. With the progress of various theoretical models, {\it{ab initio}} methods to understand the physics in strong laser fields, has made it easy to understand dynamics of collisional processes. When laser parameters like, field strength and pulse duration are made comparable with collisional interaction parameters better insight into reaction dynamics is achieved due to  coherence effects taking place during collision process [\ref{vpras}].
Laser assisted molecular excitations in ion-molecule collisions are important tool for understanding various molecular properties in depth [\ref{vpras1}-\ref{vpras3}].
Vibrational collisional excitations of $He^{+}+ CO$, system in presence of  IR laser  pulse have been studied theoretically [\ref{ADN}] .

Laser induced collisional process occurs only, if two kind of mechanism namely, collision and radiative interaction present simultaneously. Application of the laser field  during collision of two  atoms  induce population transfer from one atom to another atom under consideration. Such collisional processes proceed under joint influence of dipole-dipole, quadrupole-dipole, quadrupole-quadrupole interactions between atom and laser fields   [\ref{Lu11},\ref{Lu1},\ref{Lu22}].

Laser induced  atom-atom collision in presence of strong and weak laser fields is studied by energy transfer mechanism. Weak laser field causes inter-particle transitions and  Strong field produces ac Stark splitting in intermediate and final states of the atoms, which in turn enhances population transfer [\ref{Lu}]. 
Charge transfer mechanism induced by laser during ion-atom collisions have central place in plasma physics. Several studies are done to explore its importance [\ref{ZL1},\ref{HZ},\ref{AO},\ref{FJD}]. 
Collisional rotational dynamics of NO($X^{2} \pi$)+Ar sytem has been investigated theoretically and experimentally [\ref{Br2}]. Steric effects in NO(x) systems with different collisional partners are studied and compared [\ref{Br3}]. Other studies based on collisional processes are [\ref{Ma1},\ref{Ma2}].
Rotational alignment effects are investigated experimentally  in NO(x)+Ar inelastic collisional system using hexapole electric field method for initial state and velocity-map ion imaging technique for final state selection [\ref{Br1}]. 

The  electromagnetic field assisted collisional rotational excitation of HBr molecule due to ion (proton) impact has been investigated in reference [\ref{collision}]. Laser, molecule and ion interactions can be represented by following reaction:  

\begin{equation}
H^{+}+HBr(\upsilon,J=0)+n\hbar\omega \rightarrow H^{+}+HBr(\upsilon ,J=J^{\prime}),
\end{equation}  

with, $\upsilon$ and $J$,  as  the vibrational and rotational quantum numbers, respectively.

 {\it{HBr}} molecule is taken in its ground electronic and vibrational states. To study collisional rotational dynamics , The Schr\"{o}dinger equation,  (in atomic units i.e. $e=\hbar=m_{e}=1$):
 
\begin{equation}
\iota\frac{\partial{\psi}}{\partial t}({\bf{r}},t)=[H_{0}({\bf{r}})+V_{C}({\bf{r}}, {\bf{R}},\gamma)+V_{L}({\bf{E}}, \omega)]\psi({\bf {r}},t),
\label{lc3}
\end{equation}

 is solved using the fourth-order Runge-Kutta method. Also in eqn.(\ref{lc3}),  the term, $H_{0}$ represents field-free  Hamiltonian of the molecule and $V_{C}$, the interaction term  for   ion-molecule interaction  is:
 
\begin{eqnarray}
V_{C}({\bf{r}},{\bf{R}},\gamma)&=&-\frac{ {\bf{\mu}}({\bf{r}})\cdot {\bf{R}}}{R^{3}} \nonumber \\    
			       &=&-\frac{\mu R cos\chi}{R^{3}} \nonumber \\ 
			       &=&-\frac{\mu vt}{(b^{2}+v^{2}t^{2})^{3/2}}	,		
\label{lc4}
\end{eqnarray}

Here, $\mu(\bf{r})$ is molecule's the dipole moment  , $cos \chi = \frac{vt}{R}$ and $\bf{R}$ is defined as: 
          
\begin{equation}
{\bf{R}}(t)= {\bf{b}}+ {\bf{v}}t; {\hspace{3.5cm}} {\bf{b}}. {\bf{v}}=0 ,
\label{lc5}
\end{equation}

In above equation , $\bf{R(t)}$ represents  the position vector of the ion  with  reference to the center of mass of the molecule. Also, \lq ${\bf{v}}$\rq,\lq ${\bf{b}}$\rq  are  the relative collisional velocity and  the impact parameter, respectively. Also, $\gamma$ is the angle between $\bf{r}$ and $\bf{R}$. However, laser- molecule interaction,   $V_{L}(\bf{E},\omega)$  in the equation (\ref{lc3}) is:
 
\begin{eqnarray}
V_{L}({\bf{E}}, \omega)&=&-{\bf{\mu}}({\bf{r}}).{\bf{E(t)}} \nonumber \\
		     &=&-\mu E(t)\cos\theta \cos\omega t,	
\label{lc6}
\end{eqnarray}

Here, $\theta$ stands for  the angle between the polarization vector (of the laser pulse) and molecular axis. Ion-molecule collision and molecule-laser pulse interaction occurs  at time $t=0$,\, i.e., there is no delay time between two processes and both of them supposed to occur simultaneously. 
 
Influence of changing various parametres (molecular as well as laser) involved is explored  on collisional  rotational excitation  dynamics.  
The ion-molecule interaction term $V_{C}$ (represented by eqn.(\ref{lc4})) can be adjusted by changing impact parameter(${\bf{b}}$) and collisional velocity (${\bf{v}}$) to get desired collisional pulse for application (see figure(2) in reference [\ref{collision}). An increase in impact parameter makes peak of the asymmetric pulse broad, as a result force applied on the target molecule in kick mechanism is reduced. However, as collisional velocity is decreased area of the collision pulse increases, which indicates an increase in the energy of the pulse. Hence, one can control rotational population of the system under consideration by choosing proper parameters in the collisional pulse. 
 
Impact of collisional pulse and laser pulse  on  time evolutions of rotational probabilities of first ten rotational states ($J=0,1,2,3,4,5,6,7,8,9$)  is shown in figure(3) of [\ref{collision}].   Interaction parameters like; field strength,collisional velocity and pulse duration are $10^{-2}\;a.u.$, $0.01\;a.u.$ and $0.2\;ps$, respectively. When only Laser pulse interacts with the molecule  high rotational states  (J=6,8) are more populated than other rotational states. However when collisional pulse is applied along with laser pulse ($Sin^{2}$ HCP), the population of these states decreases and rotational states ($J=5,6$) becomes more populated. Hence it seems that collisional pulse acts like a distracting factor for rotational excitation to higher states in higher fields. However, by changing impact parameter ${\bf{b}}$, rotational excitation dynamics of the higher rotational states can be controlled. As ${\bf{b}}$ is increased, ion impact on molecules decreases and laser field becomes dominating factor in controlling the population in higher rotational levels. However, at lower values of the impact parameter ${\bf{b}}$ , ion impact on molecules is large as compare to laser field strength.
   
 Variation of  transition probabilities with the impact parameter  in lower  and high laser  field are compared in figure(6) of reference [\ref{collision}]. When field strength is small, transition probabilities of higher rotational states are small  and on increasing the field strength,  the transition probabilities of higher rotational states  increases. However, for all individual rotational states transition probabilities initially increases with impact parameter then it attains the maximum value and then it reduces to zero and it is true for both low and high field values.

As pulse proceeds with time, the extent of rotational excitation $\langle J^{2} \rangle$ shows interesting behaviour as described in figure 12(b) of reference [\ref{collision}]. It increases suddenly as the pulse proceeds and after attaining a maximum value, it decreases rapidly and attains a constant value as the laser  pulse proceeds.  However, as impact parameter is increased measure of rotational excitation increases, it is seen that peak of  $\langle J^{2} \rangle$ shifts towards right and starts diminishing with further increase in impact parameter. However,  $\langle J_{z} \rangle$, measures the orientation and it  also shows interesting behaviour with impact parameter (figure(12 b) in reference [\ref{collision}]). Maximum of  $\langle J_{z} \rangle$ increases for lower values of ${\bf{b}}$, impact parameter. As the impact parameter increases, along with primary revivals tertiary and secondary revivals also occurs for the increasing value of the impact parameters.This indicates that, as impact parameter increases the molecule enters into the transient mode.

\section{Applications of Orientation and Alignment}

Orientation are basically important  in photofragment analysis [\ref{Aub}],for  laser focusing techniques of molecular beams [\ref{Sman},\ref{Corkum}], for isotope separation [\ref{Mies}],for laser-induced isomerization [\ref{mura}],  catalysis [\ref{olte}] surface processing [\ref{Sman}] and for molecular trapping [\ref{Sman},\ref{abek}].  

In molecular alignment rotational wave-packets with high angular distributions occur.  Presence of  of rotational wave packets in gas molecule was  observed first time by Heritage etal.[\ref{JPH}]. After this discovery, many spectroscopic applications evolved [\ref{JSB},\ref{MWL},\ref{HMP}].Strong field ionization  based experiments were performed after realising the importance of aligning molecules [\ref{IVK}-\ref{DP}]. The rotational wave packets were realised to generate high degree of impulsive field free alignment[\ref{FEH}].
Adiabatic and nonadibatic alignment have been used to study high harmonic generation [\ref{RNM}-\ref{JDJM}].

Molecular alignment has also been utilised in structural measurements using x-ray diffraction [[\ref{JKO}-\ref{MOP}]. In laser induced diffraction experiments, electron's diffraction diffraction from the molecules is influenced by alignment [\ref{SNSI}].   
Experiments involving gas or liquid molecules uses controlled alignment as a basic tool for further investigation [\ref{SRT}].  Two pulse alignment technique is one of the efficient technique to enhance the alignment in field free conditions which is  further  utilized in experiment to explore the  desired system [\ref{VGE},\ref{FMJJ}].  
 Aligned rotational wave-packets are important to control many applications in optical fibres which occurs at ultra-fast time scales [\ref{NG}-\ref{VKM}]. Experimental studies mentioned in references [\ref{KMS}-\ref{XX}] also based on  the concept of aligned rotational wave packets.
Impulsive aligned structures when combined with high spatial resolved laser pulses are used in creating time dependent, complex structured photonic crystals [\ref{JDRD}].

Alignment has crucial applications in chemical reaction dynamics and stereochemistry related to molecular structures[\ref{JIJ},\ref{RT07}]. Strongly aligned structures have important role in quantum computing [\ref{EASI}] and nano-scale design [\ref{TSEI}]. It is also possible to get images of aligned molecules by a recently developed technique called laser induced electron diffraction[\ref{BWOL}].  
In molecular imaging techniques,  we can see that, when laser light interacts with the molecule, what happens to the movements of the molecule at molecular level. This technique is helpful in controlling and diagnosis of many diseases. 
Aligned molecules  have been used to generate high harmonic generations(HHGs) [\ref{NHR}], which are further helpful to get molecular images [\ref{ATLRR}].  Other imaging techniques  based studies are [\ref{PMKRA}-\ref{CZBOW}]. ,where molecule images of aligned molecules play crucial role. 
In intense laser field atoms shows non linear behaviour and they emit coherent radiations of multiple frequencies. This phenomenon is termed as  high-harmonic generation (HHG)[\ref{HHG1}]. In HHGs high energy photons can be produced  using KrF excimer lasers which have high frequency[\ref{HHG3}. Some experimental studies  for generation of HHGs are  [\ref{HHG4},\ref{HHG5}]. It is noticed in these studies that harmonic spectrum obtained has universal shape, it diminish for first few harmonics followed by plateau like behaviour when all harmonics have same strength and finally a sharp cutoff is obtained. The cut-off location is associted with optimum limit of the highest frequency. Krause etal have the maximum energy behaviour in harmonic spectrum at the end of the plateau. However Kulander etal [\ref{HHG6}] have presented well cut-off behaviour in harmonic spectrum [\ref{HHG8}].  

Molecular alignment based selective approaches like; manipulation of selective components in isotropic mixtures [\ref{SYP}] and spin selective alignment in isomers[\ref{SYPr}], rotational excitations within selective states  [\ref{SJW}], isotope selective  ionization [\ref{HTTR}] have been proposed in literature. 
Several scattering processes have been controlled through molecular alignment[\ref{EGer}-\ref{LYJH}]. In these processes molecular alignment alters dipole forces through  molecular polarzabilities[\ref{SMPF}].
 Dynamics  of molecular clusters [\ref{GGI}] and absorption spectra in transient states of   other complex molecules [\ref{Baekhoj}] has been explained  by using the concept of  aligned rotational wavepackets. 
 
The arrival of self mode locked Ti-sapphire and  chirped laser pulse have caused developments of  attosecond-femtosecond scale phenomenon related to molecular dynamics field. Through the advancement of this technology it has become possible to observe the actions of atoms and molecules at atomic level. This has opened new gates for controlling alignment and orientation.
Designing of  nanoscale molecular switches [\ref{MGMG}], ionization of polar molecules [\ref{SYS}] and their measurements [\ref{DKF},\ref{XXie}] are also crucially dependent on aligned molecules.In collision experiments collisional relaxation has been tracked by means of field free alignment [\ref{TVCB},\ref{GEFB}].

The photon dissociation of iodobenzene [\ref{MDPo}] and $I_2$ molecule [\ref{JJLa}] is also based on concept of alignment.If asymmetric top molecules are aligned in three dimensions then it is possible to photoexcite  asymmetric top molecules. Multiphoton ionizations in the strong field strongly depends on the angle between the molecular axis and filed polarization  [\ref{TSei1}-\ref{TZuo}].
When $ICl$ molecule exposed to  a $100\;fs$ laser pulse,when it is polarized parallel to alignment, upto   six electron can be easily removed by the ionizing field ,however   if the molecule is polarized perpendicular to field it is not possible to remove more than two electrons. Also in  $I^{2+}-Cl^{3+}$ channel ionization of $ICl$ is enhanced in accordance with enhanced ionization theory [[\ref{TSei1},\ref{TSei2}]]

\section{Outlook and Conclusion}
\indent We find that a HCP and a delayed zero area pulse provide an efficient mechanism for enhancing the degree of orientation of a polar molecule. We also find that the orientation and the population due to ramped pulses is enhanced in comparison to the Gaussian pulse. The results also bring in view the capability of IRL pulse in controlling the alignment dynamics of the molecule.\\ 
\indent We report on the simulations of NAREX, between the rotational states, driven by laser fields consisting of HCP and a SQP. We have shown that pulse shape have significant effect on the transition probabilities and 2DA dynamics of the molecule. We also find that the delayed pulse seems very promising in enhancing the rotational excitation  and also in controlling the molecular 2DA. The variation in  elliptic parameter also brings in view interesting and novel results of 2DA dynamics. By suitably choosing the width of delayed pulse, one can control the population of higher rotational states.  The proper choice of delay time between the two pulses decides maximum alignment along the x and z-direction.\\
\indent We also theoretically investigated the rotational excitation and alignment of molecule by train of IRL pulses of different shapes. Train of multiple weak pulses allows one to prevent unwanted destructive effects such  as ionization that usually occur with strong fields. We have studied various pulse parameters to study the rotational dynamics and alignment of molecule such as number of pulses, intensity of pulse, pulse width, shape of the pulse and pulse duration. But we find that the shape and width of the pulse are most significant in controlling the rotational dynamics and alignment of HBr molecule. \\
\indent We have also investigated the rotational states of a molecule in the presence of mixed-field tilted at an angle with respect to each other and having moderate electric field strengths. The pulse shapes taken are of SSQP and SQP type. We have shown that pulse width and pulse shape have significant effect on the population, orientation and alignment dynamics of the molecule. We also find that the delayed pulse seems very promising in enhancing the rotational excitation  and also in controlling the molecular orientation and alignment. The variation in  tilt angle also brings in view interesting and novel results of orientation and alignment dynamics. By suitably choosing the width of delayed pulse, one can control the population of higher rotational states.  The proper choice of delay time between the two pulses decides maximum orientation and alignment. The temperature also plays an important role in the orientation dynamics of the molecule. \\
\indent The above studies  prove to be quite useful in various applications in stereo dynamics, in chemical reactions, molecular separation techniques. Hence, rotational excitation remains a massive area of research in physics and chemistry. So, we believe that the theoretical results can provide an experimental basis and great potential application on the molecular alignment.\\
\section*{Acknowledgement}
One of us (VP) is thankful to  UAM, Iztapalapa, Mexico city, Mexico for hospitality during the course of this work. VP also acknowledges useful discussion  with professor Robin P. Sagar, UAM-l.


\end{document}